\documentclass[twocolumn]{aastex631}

\newcommand{\hi}{\text{H\,\sc{i}}}
\usepackage{graphicx}
\usepackage{amsmath}
\usepackage{booktabs}
\begin{document}

\title{Are Gas-rich Ultra-diffuse Galaxies and Field Dwarfs Distinct?}

\author{Khadeejah Motiwala}
\affiliation{Department of Physics, Engineering Physics and Astronomy, Queen’s University Kingston, ON K7L 3N6, Canada}

\author{Ananthan Karunakaran}
\affiliation{Dunlap Institute for Astronomy and Astrophysics, University of Toronto, 50 St. George St, Toronto, ON M5S 3H4, Canada}

\author{Kristine Spekkens}
\affiliation{Department of Physics, Engineering Physics and Astronomy, Queen’s University Kingston, ON K7L 3N6, Canada}

\author{Nikhil Arora}
\affiliation{Department of Physics, Engineering Physics and Astronomy, Queen’s University Kingston, ON K7L 3N6, Canada}
\affiliation{Arthur B. McDonald Canadian Astroparticle Research Institute,  Queen’s University, Kingston, ON K7L 3N6, Canada}

\author{Arianna Di Cintio}
\affiliation{Universidad de La Laguna, Avda. Astrofísico Fco. Sánchez, 38205, La Laguna Tenerife, Spain}
\affiliation{
Instituto de Astrofísica de Canarias, Calle Via Láctea s/n, 38206, La Laguna, Tenerife, Spain}

\author{Anna C. Wright}
\affiliation{Center for Astrophysical Sciences, William H. Miller III Department of Physics \& Astronomy, Johns Hopkins University, 3400 N. Charles Street, Baltimore, MD
21218, USA}
\affiliation{Center for Computational Astrophysics, Flatiron Institute, 162 Fifth Avenue, New York, NY 10010, USA}

\author{Dennis Zaritsky}
\affiliation{Steward Observatory, University of Arizona, 933 North Cherry Avenue, Tucson, AZ 85721-0065, USA}

\author{Andrea V. Macci\`o}
\affiliation{New York University Abu Dhabi, PO Box 129188, Abu Dhabi, United Arab Emirates}
\affiliation{Center for Astrophysics and Space Science, New York University Abu Dhabi, Abu Dhabi, PO Box 129188, Abu Dhabi, UAE}
\affiliation{Max-Planck-Institut für Astronomie, Königstuhl 17, D-69117~Heidelberg, Germany}

\begin{abstract}

We explore the differences in gas-rich field Ultra-Diffuse Galaxies (UDGs) and classical dwarf galaxies using an extensive atomic gas (HI) follow-up survey of optically-selected UDG candidates from the Systematically Measuring Ultra-Diffuse Galaxies (SMUDGes) catalogue. We also compare the SMUDGes-HI observations with two state-of-the-art cosmological hydrodynamical simulations: Numerical Investigation of a Hundred Astrophysical Objects (\textsc{nihao}), where UDGs form through a series of bursty star formation episodes and \textsc{Romulus25}, in which UDGs occupy dark matter halos with high spins as a result of major mergers. Although the suggested formation scenarios for UDGs within these simulations are different, the present-day HI masses $M_\mathrm{HI}$, stellar masses $M_*$, and star formation rates $SFR$ of simulated galaxies are qualitatively and quantitatively consistent with each other and with the observed SMUDGes-HI sample. We find that when controlling for $M_*$, there is a positive correlation between the gas richness $M_\mathrm{HI}/M_*$ and the effective optical radius $R_{eff}$, and that this trend is not different between the UDG and dwarf populations, within the measured scatter. Taken together, our results suggest that gas-rich, star-forming UDGs and dwarfs are not distinct galaxy populations, either observationally or in simulations.

\end{abstract}

\keywords{galaxies: dwarf, galaxies: formation, galaxies: evolution}

\section{Introduction}
In the standard cold dark matter ($\Lambda$CDM) framework, galaxies form hierarchically, with the most massive galaxies embedded in extensive dark matter haloes that were formed by subsequent mergers of smaller haloes \citep{white1978core, 1991ApJ...379...52W,1992ApJ...390L..53K, kauffmann1993formation}. Dwarf galaxies ($M_* \leq \,\sim\!10^9~\mathrm{M}_\odot$) reside in these low-mass haloes and are the most numerous galaxy systems in the Universe \citep{1976ApJ...203..297S}. However, despite their abundance, their relatively low stellar masses and small sizes mean they have historically been difficult to observe. 

The faint end of the galaxy luminosity function is dominated by low surface brightness galaxies (LSBs) which have traditionally been defined as galaxies fainter than the limiting isophote of the night sky \citep{impey1988virgo, bothun1991extremely}. Currently, they are often defined as galaxies with a surface brightness that falls below the detection limit of the Sloan Digital Sky Survey ($\sim24$ mag arcsec$^{-2}$ in the $g$-band \citealt{2000AJ....120.1579Y}). 

In recent years, a new generation of small aperture optical telescopes (e.g. \citealt{abraham2014ultra, javanmardi2016dgsat, 2018PASJ...70S...4A}) and image-searching techniques (e.g. \citealt{2017ApJ...850..109B, 2018A&A...615A.105M, 2018MNRAS.478..667P, 2019ApJS..240....1Z}) have renewed interest in LSBs. Notably, with the use of deep imaging with the Dragonfly Telephoto Array, \citet{van2015forty} uncovered 40 extended LSBs in the Coma Cluster, and called them Ultra-Diffuse Galaxies (UDGs). The authors also proposed a definition based on central surface brightness ($\mu_{\mathrm{0,g}}\geq 24$ mag arcsec$^{-2}$) and size (effective radius $R_{\mathrm{eff}}\geq1.5$ kpc) for these objects, which has since been widely adopted in the field. 

With stellar masses typical of dwarf galaxies but physical sizes comparable to those of Milky Way-type galaxies, UDGs have garnered significant interest. Since their discovery, thousands of UDGs have been found in cluster environments (e.g. \citealt{2015ApJ...807L...2K, mihos2015galaxies, van2016abundance, venhola2017fornax}), as well as in lower density groups (e.g. \citealt{crnojevic2016extended, 2016ApJ...833..168M, 2016ApJ...830L..21T, 2018ApJ...855...28S}) and the field (e.g. \citealt{2016AJ....151...96M, bellazzini2017redshift, rong2017universe, 2018ApJ...857..104G}). UDGs in high density environments are generally observed to be redder than their isolated counterparts, which are predominantly blue with regions of denser star formation \citep{roman2017ultra, 2019MNRAS.488.2143P}.

The extreme properties of UDGs along with their sheer abundance across these diverse environments have sparked interest in their formation and evolution. For UDGs in clusters, external effects, such as pressure stripping of the ram and tidal forces, could be responsible for quenching \citep{2015MNRAS.452..937Y, janowiecki2019environment, tremmel2020formation, moreno2022galaxies}. \citet{van2015forty} propose that UDGs form in Milky Way mass haloes but lose a significant portion of their gas at early times, hindering star formation and producing ``failed" $L_*$ galaxies. 

On the other hand, UDGs in the field may form through internal mechanisms. For example, \citet{2016MNRAS.459L..51A} suggest that UDGs are ``genuine dwarfs" that form in dwarf dark matter haloes with higher-than-average angular momenta. A number of UDGs with low halo masses have been found to support this scenario (e.g. \citealt{trujillo2017nearest, kovacs2019constraining}). \citet{di2017nihao} show that field UDGs in the Numerical Investigation of a Hundred Astrophysical Objects (\textsc{nihao}) cosmological simulations form through periods of repeated star formation. On the other hand, \citet{wright2021formation} demonstrate that field UDGs in the \textsc{Romulus25} simulations underwent high redshift major mergers that increased their specific angular momenta, causing star formation to migrate from the centers of the galaxies to the outskirts. 

This diversity of proposed formation mechanisms fuels the question: are UDGs a distinct population of galaxies or simply the large and faint extension to the dwarf galaxy population? Indeed, the UDGs in the \textsc{nihao} \citep{di2017nihao} and \textsc{Romulus25} \citep{wright2021formation} simulations are not formed from different physics, but rather as a result of more intense periods of star formation and more prograde mergers, respectively. Likewise, \citet{van2015forty} clarify that the term ``UDG" does not necessarily mean that they are ``distinct from the rest of the dwarf galaxy population". Moreover, studies of UDGs in clusters find that they are not different from dwarfs: for example, \citet{2016ApJ...819L..20B} and \citet{Conselice_2018} hypothesize that UDGs display similar properties to quenched dwarfs, while \citet{2023MNRAS.519.1545C} find that the two populations have similar metallicity gradients, \citet{2024ApJS..271...52Z} find that they have the same structural parameters (but see also \citealt{van2018galaxy, forbes2020globular}), and \citet{Buzzo25} find that they have similar globular cluster populations.

In order to constrain formation models, large samples of UDGs  are required with distance information in order to confirm their large physical sizes - the estimation of which is challenging in the case of such faint objects. In cluster environments, distances can be reliably inferred by their projected proximity to brighter objects within the cluster. In the field, this is not possible. Optical spectroscopic distances to some field UDGs have been obtained (e.g., \citealt{van2015spectroscopic, 2019MNRAS.484.3425M, 2019ApJ...884...79C, 2021ApJ...923..257K}), but on a larger survey scale, the requisite integration times become prohibitive. 

Alternatively, the neutral Hydrogen (HI) spectral line emitted by gas-rich field UDGs can provide an efficient measure of distance and the HI content. \citet{leisman2017almost} catalog 115 gas-rich UDGs with distance information from the Arecibo
Legacy Fast ALFA (Arecibo L-band Feed Array) extragalactic untargetted extragalactic HI survey (e.g. \citealt{giovanelli2005arecibo, haynes2011arecibo}). HI follow-ups of optically detected UDGs have also been fruitful (e.g. \citealt{papastergis2017hi, 2018ApJ...855...28S, karunakaran2020systematically, karunakaran2024}). 

The Systematically Measuring Ultra-Diffuse Galaxies survey (SMUDGes; \citealt{zaritsky2018systematically,zaritsky2021systematically,zaritsky2022systematically,2023ApJS..267...27Z}) identifies UDG candidates in optical images from the Dark Energy Spectroscopic Instrument's (DESI) Legacy Surveys \citep{dey2019overview}. The SMUDGes-HI program followed up 378 of the candidates in HI -- the largest UDG HI follow-up effort to date -- using the Robert C. Byrd Green Bank Telescope (GBT). SMUDGes-HI detected 110 gas-rich UDGs and foreground dwarfs, distinguishing between the two using the distances implied by the recessional velocity of the HI spectrum \citep{karunakaran2020systematically, karunakaran2024}. 

In this work, we compare gas-rich UDGs and dwarfs in the SMUDGes catalogue, which were  all selected as UDG candidates from optical images with the same criteria \citep{2023ApJS..267...27Z}. This makes SMUDGes especially well-suited to a comparison study of this nature. We test whether confirmed UDGs and dwarfs from SMUDGes populate distinct regions in parameter space. Moreover, we compare the SMUDGes observations with UDGs and dwarfs in the \textsc{NIHAO} and \textsc{Romulus25} simulations, which are well-suited as a pair because their hydrodynamics-solving codes are very similar. 
We make a series of comparisons between the properties of UDGs, dwarfs, and simulations to look for differences between observed and simulated samples that might signal different formation pathways.

In Section \ref{sec:data}, we describe the selection criteria that we impose on the parent samples to compose our final comparison samples. We present the results of our comparison study in Section \ref{sec:results} and discuss the results in Section \ref{sec:discussion}. We conclude and summarize our findings in Section \ref{sec:conclusion}.

\section{Constructing Comparison Samples} \label{sec:data}

\subsection{Parent SMUDGes Sample}

We use the sample of HI-confirmed UDGs and dwarfs from the SMUDGes-HI survey \citep{karunakaran2024} as our starting point. 
SMUDGes-HI follows up 378 optically-selected galaxies in SMUDGes with deep HI observations from the GBT. 
The photometric properties of the targets are derived from Legacy Survey imaging \citep{dey2019overview}, are modeled as exponentials using \textsc{Galfit} \citep{2010AJ....139.2097P} and then selected on $g$-band central surface-brightness ($\mu_{\mathrm{0,g}} > 23.5$ mag arcsec$^{-2}$) and angular half-light radius ($r_{\mathrm{e}} \geq  5.3$ arcsec, which corresponds to $R_{\mathrm{eff}}\geq 2.5$ kpc at the distance of the Coma Cluster; \citealt{2023ApJS..267...27Z}).

SMUDGes-HI targets were selected using a $g$-band magnitude cut $m_g \leq 19$ mag to ensure HI detections or meaningful upper limits from the GBT with  integrations times of up to a few hours \citep{karunakaran2024}. Additionally, in later observing campaigns, candidates were required to have discernible GALEX (Galaxy Evolution Explorer; \citealt{2007ApJS..173..185G}) UV emission in order to maximize the number of HI detections. Star formation rates are determined from GALEX photometry using the relations from \citet{2006ApJS..164...38I}. 

A number of quantities can be derived from HI spectra for the detections that make up the SMUDGes-HI sample \citep{karunakaran2024}. Crucially, the redshift of the line gives distance $D$ from which the physical sizes of the UDG candidates are inferred. They are then classified as bona fide UDGs using the \citet{van2015forty} definition ($\mu_{\mathrm{0,g}} > 24$ mag arcsec$^{-2}$ and $R_{\mathrm{eff}}> 1.5$ kpc); UDG candidates for which these criteria are not met are classified as dwarfs. The distances are also used to compute HI masses from the HI line integral, stellar masses from the $m_g$ and the $g-r$ color using the color - stellar mass relations of \citet{2017ApJS..233...13Z}, and star formation rates from GALEX photometry using the relations from \citet{2006ApJS..164...38I}. We use the $R_{\mathrm{eff}}$, HI masses ($M_\mathrm{HI}$), stellar masses ($M_{*}$), and star formation rates ($SFR$) and their uncertainties as reported by \citet{karunakaran2024} for the SMUDGes-HI sample in this work. 

\subsection{Parent Simulation Samples}

We draw simulated dwarfs and UDGs for comparison with the SMUDGes-HI sample from \textsc{nihao} \citep{Blank2019, Maccio2020} and \textsc{Romulus25} \citep{2017MNRAS.470.1121T}: they are complementary since their hydrodynamics solvers are very similar, but \textsc{nihao} is a zoom simulation while \textsc{Romulus25} is a volume simulation. Table~\ref{tab:sims} compares their basic properties that are relevant to our analysis. In what follows, we briefly describe the two simulations here and the selection criteria for dwarfs and UDGs for our comparison. 

%
\begin{table}[]
\centering
\begin{tabular}{@{}cccc@{}}
\toprule
Property              & Unit                & \textsc{nihao}                            & \textsc{Romulus25}                                   \\ 
(1)                   & (2)                 & (3)                               & (4)                                          \\ \midrule
Type                  &                     & Zoom-in                           & Volume                                       \\
Hydrodynamics         &                     & \textsc{Gasoline2.0}              & \textsc{ChaNGa}                              \\
IMF                   &                     & Chabrier                          & Kroupa                                       \\
Box-size              & ${\rm Mpc\,h^{-1}}$ & 60                                & 25                                           \\
$m_{\rm dark}$        & M$_{\rm \odot}$     & $2.1\times 10^{5}$                & $3.39\times 10^{5}$                          \\
$\epsilon_{\rm dark}$ & pc                  & 465                               & 350                                          \\
$\epsilon_{\rm gas}$  & pc                  & 199                               & 350                                           \\
$n_{\rm SF}$          & cm$^{-3}$           & 10.3                              & 0.2                                          \\
$T_{\rm SF}$          & K                   & $1.5 \times 10^4$                             & $1 \times 10^4$                                        \\

\bottomrule
\end{tabular}
\caption{A summary of \textsc{nihao} and \textsc{Romulus25} simulation properties, from \citet{2017MNRAS.470.1121T} and \citet{Blank2019} respectively: 
dark matter particle mass ($m_{\rm dark}$), softening lengths for dark matter ($\epsilon_{\rm dark}$) and gas ($\epsilon_{\rm gas}$), star formation density ($n_{\rm SF}$) and temperature ($T_{\rm SF}$) thresholds, and the adopted Initial Mass Function (IMF), and the adopted hydrodynamics solver.
\label{tab:sims}
}
\end{table}

\subsubsection{NIHAO Parent Sample}

\textsc{nihao} \citep{wang2015nihao, Blank2019, Maccio2020} are a set of $\sim$130 zoom-in cosmological simulations in a flat $\Lambda\rm{CDM}$ cosmology \citep{planck14} using \textsc{Gasoline2.0} \citep{Wadsley2017} Tree smoothed particle hydrodynamics (TreeSPH) code 
to produce galaxies with stellar masses ranging from $10^{6}-10^{12}\,{\rm M_{\odot}}$ at $z=0$. 
All simulations have similar resolution, containing $\sim 10^6$ dark matter particles with a softening length of $\epsilon_{\rm dark} = 465\,{\rm pc}$ and mass resolution of $2.1\times 10^5\,{\rm M_{\odot}}$.
Each simulation also contains $\sim 10^6$ gas particles, with a softening length of $\epsilon_{\rm gas} = 199\,{\rm pc}$ and a typical particle mass of $3.1\times 10^4\,{\rm M_{\odot}}$. 

NIHAO galaxies form stars according to the Kennicutt-Schmidt law \citep{Kennicutt1998} with suitable gas temperature and density thresholds, $\rm T_{\rm SF}< 1.5 \times 10^4\,K$ and $\rm n_{\rm SF}>10.3\, cm^{-3}$, and follow the Chabrier initial mass function \citep{Arora2022}. 
Massive stars with $\rm 8\,M_{\odot}<M_{star}<40\,M_{\odot}$ ionize the interstellar medium (ISM) before their supernova (SN) explosions~\citep{Stinson2006, wang2015nihao}. 
This ``early stellar feedback" (ESF) mode is set to inject 13 per cent of the total stellar flux of $2\times 10^{50}\,{\rm erg\,M_{\odot}^{-1}}$ into the ISM through blast-wave SN feedback that injects both energy and metals into the ISM.
This energy is injected into high density gas and radiated away due to efficient numerical cooling \citep{Stinson2006}. 
For gas particles inside the blast radius, cooling is delayed by $30\,{\rm Myr}$ \citep{Stinson2013}.
The stellar feedback does not have any variability with halo mass and/or redshift.

\textsc{nihao} also includes subgrid models for turbulent mixing of metals and energy \citep{keller2014superbubble}, UV heating, ionization, and metal cooling \citep{shen2010enrichment} and super massive black hole (SMBH) growth and feedback \citep{blank2019nihao, Maccio2020}.
NIHAO simulations have already proven successful at matching various observational aspects of galaxy formation and evolution 
\citep{Dutton2007, Maccio2016, di2017nihao, dicintio2019, Blank2021, Arora2022, arora2023manga}.

Galaxies in \textsc{nihao} are isolated centrals, their masses range from Milky Way-types to dwarfs and they have a wide array of merger histories, concentrations, and spins. UDGs in \textsc{nihao} live in dwarf-mass dark matter halos ($M_{\mathrm{halo}} \leq 10^{10.53}~\mathrm{M}_\odot$) and undergo periods of intense star formation which drive outflows \citep{di2017nihao}. The dark matter halo expands and redistributes the stars to larger radii resulting in a larger, more diffuse galaxy.

We select dwarfs and UDGs from simulated \textsc{nihao} galaxies with $\log_{10}(M_*/M_{\odot})\leq 9.5$ at $z=0$.
Stellar masses are measured using all star particles within $0.2R_{\rm 200}$ and star formations rates are measured over the last 100\,Myr.
To distinguish UDGs from dwarfs, central surface brightnesses ($\mu_{0,g}$) and effective radii ($R_{\mathrm{eff}}$) are needed.
Using the \texttt{pynbody} package \citep{Pontzen2013}, we calculate the face-on SDSS-\textit{g} band surface brightness profiles for the stellar distribution.
$R_{\mathrm{eff}}$ is then defined as the radius enclosing half the total light (the integral of the profile), and $\mu_{0,g}$ is the median surface brightness for the three innermost points. 
The \hi\, fraction for each gas particle is calculated using the radiative transfer prescription from \citet{2013MNRAS.430.2427R}, and \hi\, masses are calculated using all gas particles enclosed within $4 R_{\mathrm{eff}}$.

\subsubsection{Romulus25 Parent Sample}

\textsc{Romulus25} \citep{2017MNRAS.470.1121T} is a high-resolution cosmological simulation that performs hydrodynamics with the parallel N-body + SPH code \textsc{ChaNGa} \citep{Menon2015}, which is built on \textsc{Gasoline} \citep{wadsley2004gasoline}. 
In this way, the hydrodynamics solvers in \textsc{nihao} and \textsc{Romulus25} are similar with the only difference being that the \textsc{Romulus25} hydrodynamics includes a gravity improvement that speeds up N-body calculations \citep{Wadsley2017}. \textsc{Romulus25} consists of a 25\,Mpc co-moving box evolved to $z=0$. Dark matter particles are oversampled relative to gas by a factor of 3.375; this allows for similar masses of dark matter ($3.39 \times 10^5~\rm{M}_\odot$) and gas ($2.12 \times 10^5~\rm{M}_\odot$) particles, thereby reducing numerical effects from two-body interactions and  enabling better tracking of SMBH dynamics \citep{2015MNRAS.451.1868T}.

In \textsc{Romulus25}, stars form stochastically from gas particles that are sufficiently dense ($n_{\rm SF}>0.2~\rm{cm}^{-3}$) and cold ($T_{\rm SF}<10^4~\rm{K}$), following a \citet{kroupa2001variation} initial mass function. The resulting star particle has mass corresponding to 30 percent of the mass of the original gas particle. As in \textsc{nihao}, massive stars explode in Type II SNe and deposit energy into the surrounding gas via the \citet{stinson2006star} feedback prescription. Additionally, lower mass stars contribute to feedback via Type Ia SNe and stellar winds \citep{1994ApJ...435...22K}, while mass and metals are mixed into the ISM following \citet{shen2010enrichment} and \citet{2015MNRAS.448..792G}. \textsc{Romulus25} implements the novel SMBH prescription in \citet{tremmel2019introducing}, in which seed SMBHs form in low-metallicity ($Z<3\times10^{-4}\rm{Z}_\odot$), high-density ($n>3~\rm{cm}^{-3}$) gas that is still too hot to form stars. This ensures SMBHs emerge in rapidly collapsing regions, most often within the first 1 Gyr of the simulation. SMBH orbits are tracked with the dynamical friction model given by \citet{2015MNRAS.451.1868T}. \textsc{Romulus25} produces realistic galaxies across its full range of resolved masses and accurately replicates observed high-redshift star formation and SMBH growth \citep{2017MNRAS.470.1121T}.

For \textsc{Romulus25}, our parent sample builds on the dwarf and UDG sample presented in \citet{wright2021formation} which spans the stellar mass range $10^{7.2} < M_*/\mathrm{M}_\odot < 10^{8.7}$, and also includes additional objects with stellar masses up to $M_* \simeq 10^{9.25}~\mathrm{M}_\odot$ at $z=0$. S\'ersic profiles are fit to the $z=0$ $g$-band surface brightness
profile of each galaxy, from which the central surface brightness $\mu_{0,g}$ and half-light radius $r_{\rm{eff}}$ are calculated and used to identify UDGs following the \citet{van2015forty} definition. Star formation in \textsc{Romulus25} is calculated over the last 25-250 Myr of the simulation. HI masses are calculated using all gas particles within $R_{200}$. The HI fraction of each gas particle is calculated as a combination of the ionization state of the particle (based on the reionization background from \citealt{2012ApJ...746..125H}, collisional ionization rates from \citealt{1997NewA....2..181A}, etc.) and the metallicity of the particle (e.g., low temperature metal-line cooling from \citealt{2001ApJ...552..464B}).


\subsection{Observed and Simulated Comparison Samples}
\label{sec:compsample}

From the parent samples described above, we define comparison samples with properties that are sufficiently well-matched to afford a direct data-model comparison. Specifically, we ensure that the comparison samples include only the most reliable observations, and that simulations are well-matched in the range of surface brightness, effective radius, gas-richness, and specific star formation rate. A summary of the selection criteria and the corresponding sample statistics is given in Table \ref{tab:cuts}.

\begin{table*}[t]
    \centering
    \begin{tabular}{c|c|c|c}
        \hline
        \rule{0pt}{3ex} 
        \textbf{Selection criteria} & SMUDGes-HI & \textsc{Romulus25} & \textsc{nihao}  \\
        \hline
        \rule{0pt}{3ex} Parent Sample Size & 110 & 599 & 171 \\
        \rule{0pt}{3ex} $SNR_{\mathrm{NUV}} \geq 2$ & 100 & - & - \\
        \rule{0pt}{3ex} $23.5<\mu_{0,g}/\mathrm{mag~arcsec}^{-2}<26$ & 92 & 281 & 45 \\
        \rule{0pt}{3ex} 
        $sSFR \geq 10^{-11.5}~\mathrm{yr}^{-1}$ & 92 & 207 & 26 \\
        \rule{0pt}{3ex} 
        $f_{\mathrm{HI}} > 0.2$ & 92 & 196 & 26 \\
        \rule{0pt}{3ex} 
        $1.5 < R_{\mathrm{eff}}/\mathrm{kpc} < 10$ & 78 (34 UDGs, 44 dwarfs) & 132 (92, 40) & 18 (11, 7) \\
        \hline
    \end{tabular}
    \caption{Summary of selection cuts applied to the SMUDGes-HI, \textsc{Romulus25}, and \textsc{nihao} parent samples to produce the comparison samples. The cuts are applied in the order they are listed, and the numbers in each column are the galaxies that remain after the cut in that row is applied. The numbers in the bottom row constitute the comparison samples, with the breakdown between UDGs and dwarfs (defined by the surface brightness threshold $\mu_{0,g} \geq 24$ mag arcsec$^{-2}$ for UDGs) given in parentheses.}
    \label{tab:cuts}
\end{table*}

We select galaxies from the SMUDGes-HI parent sample with a sufficiently high near-UV (NUV) signal-to-noise-ratio ($SNR_{\mathrm{NUV}} \geq 2$) so that we only compare galaxies with confident star formation rates $SFR_{\mathrm{NUV}}$ to simulations. Moreover, we limit the sample to galaxies with central $g$-band surface brightness $\mu_{\mathrm{0,g}} > 23.5$ mag arcsec$^{-2}$ which corresponds to the SMUDGes selection threshold \citep{2023ApJS..267...27Z}, since some brighter targets were included in SMUDGes-HI based on preliminary surface brightness models. We also apply this surface brightness cut to the \textsc{nihao} and \textsc{Romulus25} parent UDG samples to match the brightness and sensitivity limits of SMUDGes. 

Moreover, to both simulated samples, we make a cut on specific star formation rate (sSFR = SFR/M$_* \geq 10^{-11.5}~\mathrm{yr}^{-1}$) to ensure that we select star-forming objects, as implied by the GALEX detection criterion in the observations. We also require that simulated galaxies have an HI gas fraction $f_{\rm HI} > 0.2$ to ensure that they would have been detectable in the SMUDGes-HI follow-up, since the observing times for the campaigns is set to detect gas fractions above this threshold \citep{karunakaran2024}. Our results do not depend on the precise numerical thresholds adopted for these cuts as long as they select the star-forming objects from the parent samples.

Finally, to both simulated and observed samples, we enforce a size criterion of $1.5 < R_{\rm{eff}}/\rm{kpc} < 10$. 
The lower size limit is applied as simulated low mass galaxies tend to be overquenched due to the BH seed mass being fairly high in the simulated galaxies, particularly in \textsc{Romulus25} \citep{wright2021formation}. The upper size limit is set to eliminate the few physically large \textsc{nihao} galaxies which would be hard to detect in SMUDGes by virtue of the background subtraction applied to Legacy Survey objects \citep{dey2019overview}.  

After all of the cuts described above and summarized in Table~\ref{tab:cuts} are applied, we are left with 78 galaxies from SMUDGes-HI, and 132 and 18 from \textsc{Romulus25} and \textsc{nihao} respectively, in our comparison samples. We classify UDGs in all three samples following the \citet{van2015forty} surface brightness criterion: $\mu_{0,g} \geq 24$ mag arcsec$^{-2}$. Note that since all galaxies in our comparison samples are larger than $R_{\mathrm{eff}} > 1.5$ kpc by construction, the surface brightness criterion is the only discriminating property between UDGs and dwarfs. We test a more relaxed size criterion ($R_{\mathrm{eff}}>1$ kpc) for UDGs as some authors (e.g. \citealt{di2017nihao, 2023MNRAS.519.1545C}) have done, to confirm that our results do not change significantly with a difference in UDG size definition. Following this, we are left with 34 UDGs and 44 dwarfs in the SMUDGes comparison sample, 92 UDGs and 40 dwarfs in the \textsc{Romulus25} comparison sample, and 11 UDGs and 7 dwarfs in the \textsc{nihao} comparison sample.

\begin{figure*}[t!]
\begin{center}
\includegraphics[width=0.72\textwidth]{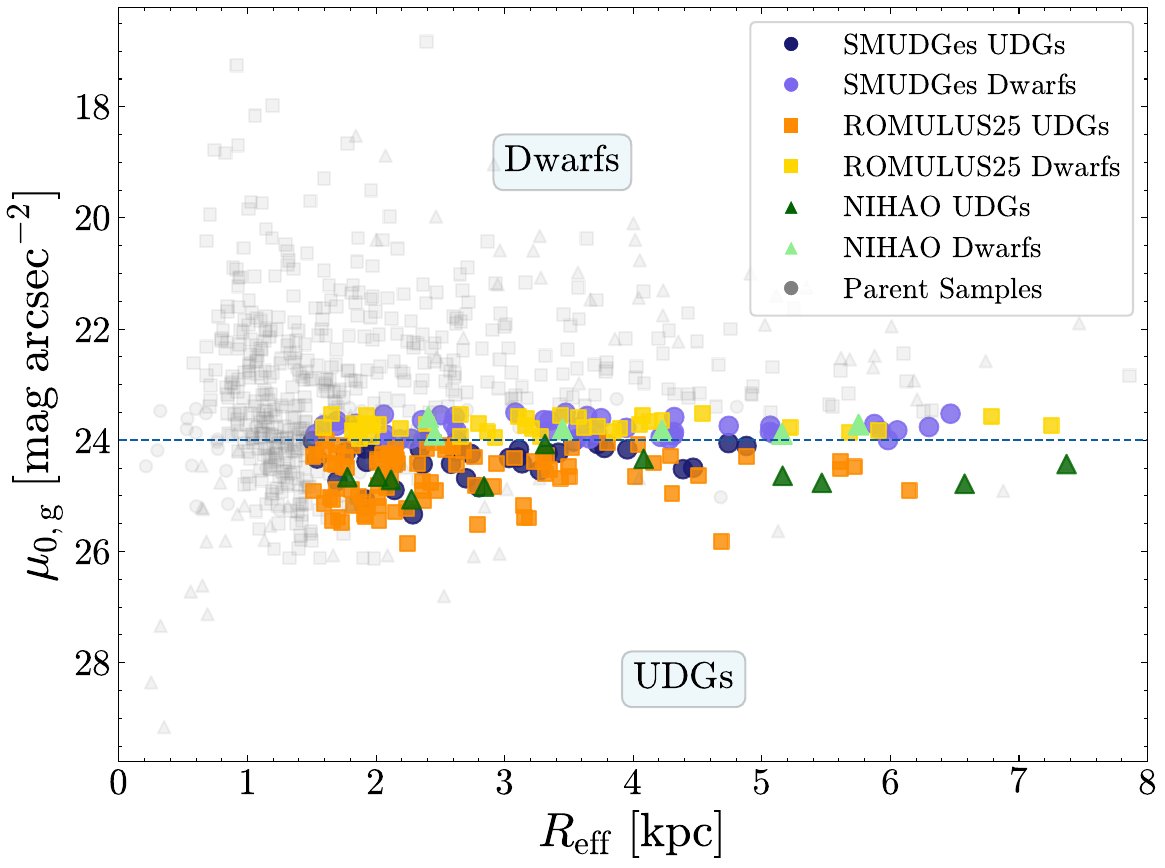}
\caption{Distribution in central surface brightness and size of the parent samples for SMUDGes-HI \citep{karunakaran2024},  \textsc{Romulus25} \citep{wright2021formation} and \textsc{nihao} \citep{blank2019nihao,Maccio2020}. The colored symbols show the comparison samples defined by the cuts summarized in Table~\ref{tab:cuts}, and the horizontal line at $\mu_{0,g} = 24$ mag arcsec$^{-2}$ represents the $g$-band central surface brightness limit that distinguishes UDGs from dwarfs. The comparison samples are the diffuse, gas-rich subset of the dwarfs and UDGs in the parent samples. 
\label{fig:sbreffful}}
\end{center}
\end{figure*}

Figure \ref{fig:sbreffful} shows the distribution of galaxies in the observed and simulated comparison samples (colored points) in relation to their parent samples (gray points) in central surface brightness - size space. By construction, the comparison samples span a smaller range in this space than their parent samples for both simulations and the observations. In particular, Figure \ref{fig:sbreffful} and Table~\ref{tab:cuts} illustrate how the surface brightness, size and gas richness cuts that form the simulated comparison samples -- which are required for direct comparisons to SMUDGes-HI observations -- produce strong differences in size and surface brightness distributions between the parent and comparison samples: the latter represent the gas-rich, star-forming, and diffuse subset of the simulated dwarfs and UDGs in \textsc{Romulus25} and \textsc{nihao}, respectively.

\begin{figure*}[hbt!]
\begin{center}
\includegraphics[width=0.72\textwidth]{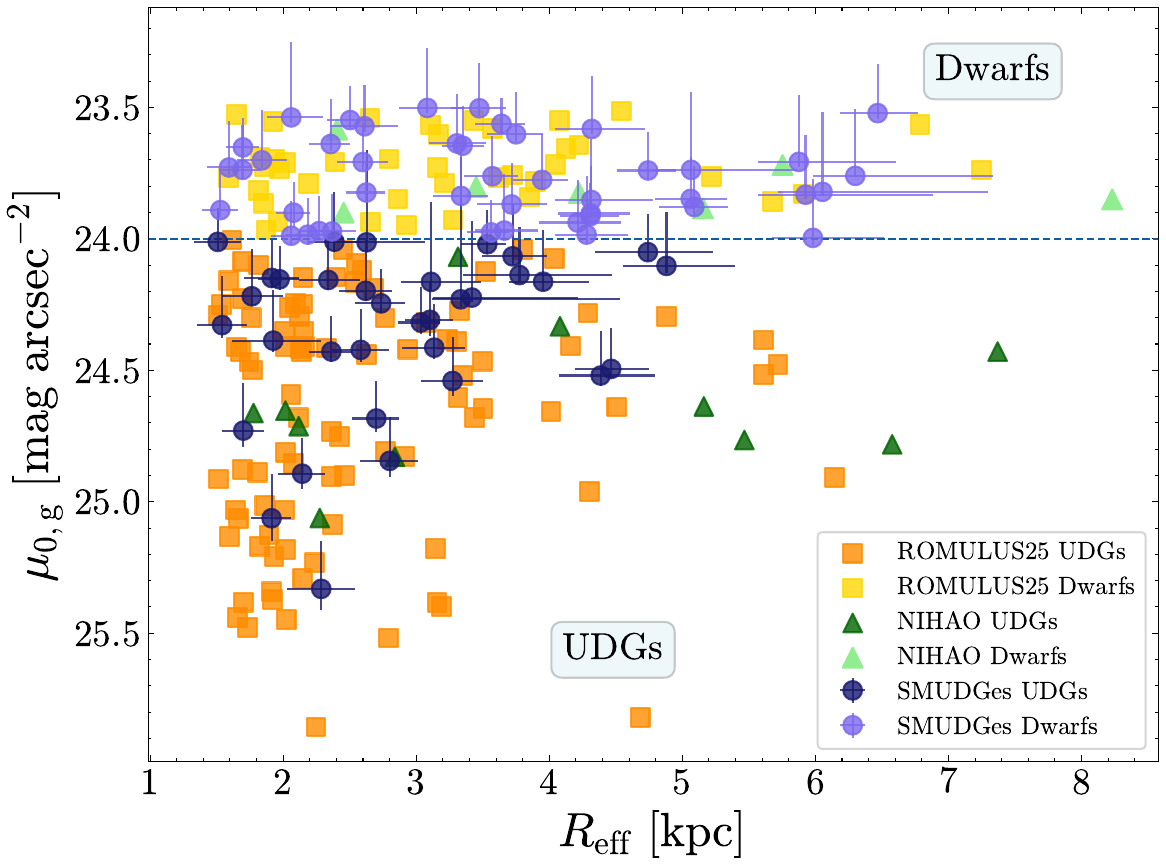}
\caption{Distribution of central surface brightness and size for the comparison samples of observed SMUDGes-HI galaxies and simulated \textsc{Romulus25} and \textsc{nihao} galaxies. The horizontal line at $\mu_{0,g} = 24$ mag arcsec$^{-2}$ represents the $g$-band central surface brightness limit of UDGs. There is strong overlap between the distributions of observed and simulated objects.
\label{fig:sbreff}}
\end{center}
\end{figure*}

Figure~\ref{fig:sbreff} presents the distribution of central surface brightnesses and sizes across the comparison samples, and including error bars that represent the uncertainties on the measured SMUDGes-HI quantities. There is strong overlap between the distributions of observed and simulated objects in this parameter space, enabling meaningful direct comparisons between observed dwarfs and UDGs as well as between observed systems and their simulated counterparts.    

\section{Results} \label{sec:results}

\begin{figure*}[hbt!]
    \centering
    \includegraphics[width=0.7\textwidth]{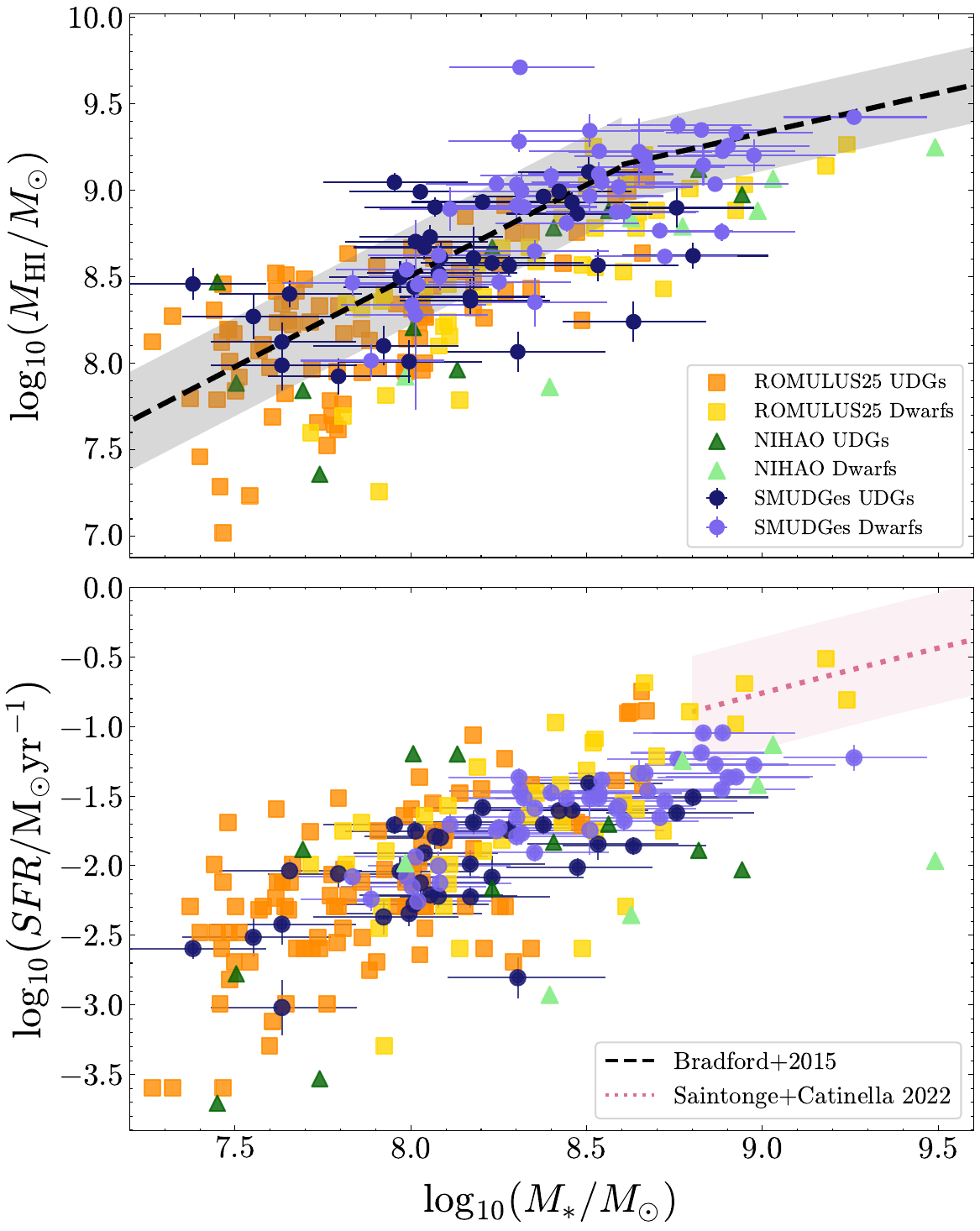}
    \caption{Distribution of comparison sample UDGs and dwarfs in $M_{\mathrm{HI}}$ (top) and $SFR$ (bottom) as a function of $M_*$. In the top panel, the black line and shaded gray region show the relation and 1$\sigma$ scatter derived by \citet{2015ApJ...809..146B} for field dwarfs. In the bottom panel, the pink line and shaded region show the star forming main sequence and 1$\sigma$ scatter from \citet{saintonge22}, plotted over the mass range of the objects in that sample.} 
    \label{fig:mhimstarsfr}
\end{figure*}


With our comparison samples defined, we proceed to examine their properties. We focus on combinations of properties that are available for both the simulated and observed samples: Section \ref{sec:res-props} compares $M_*$, $M_{HI}$ and $SFR$, while Section \ref{sec:res-gr} examines the gas richness - size ($M_{HI}/M_* - R_{eff}$) relation.

\subsection{Gas, Stars and Star Formation Rates}\label{sec:res-props}

Figure~\ref{fig:mhimstarsfr} shows the $M_{HI} - M_*$ (top) and $ SFR - M_*$ (bottom) relations for simulated and observed dwarfs and UDGs from the comparison samples defined in Section~\ref{sec:compsample}. To highlight key similarities and differences between the various distributions, we plot residuals for each sample relative to fiducial relations in Figure~\ref{fig:resplotcomb}. 

We quantify the probability that two comparison samples are drawn from the same $M_{HI} - M_*$ or $SFR - M_*$ distribution using the PQMass test \citep{lemos2024pqmass}. Briefly, the test approximates the integral of the density (probability mass) across different regions in parameter space, and does not make any assumptions about the density of the true distribution. We use 10 realizations of 100 tesselations of a space to compute the $p$-value for each test, and adopt $p<0.001$ as the threshold below with the two distributions differ significantly. We note that we do not perform statistical tests on the \textsc{NIHAO} samples given their small sizes. 

\subsubsection{$M_{\rm{HI}}-M_*$ Relation}
In the top panel of Figure \ref{fig:mhimstarsfr}, we compare the distribution of simulated and observed galaxies in $M_{HI}$ and $M_*$. Generally, both simulated and observed samples overlap in this space. There is particularly strong overlap between the SMUDGes-HI dwarfs and UDGs, and between the \textsc{Romulus25} dwarfs and UDGs: the two types of objects are well-mixed along their $M_{HI} - M_*$ relations. PQMass tests comparing SMUDGes-HI dwarfs and UDGs, and comparing \textsc{Romulus25} dwarfs and UDGs, confirm that each pair are likely drawn from the same distributions ($p = 0.06$ and $p = 0.002$, respectively).

\begin{figure*}[hbt!]
    \centering
    \includegraphics[width=0.7\textwidth]{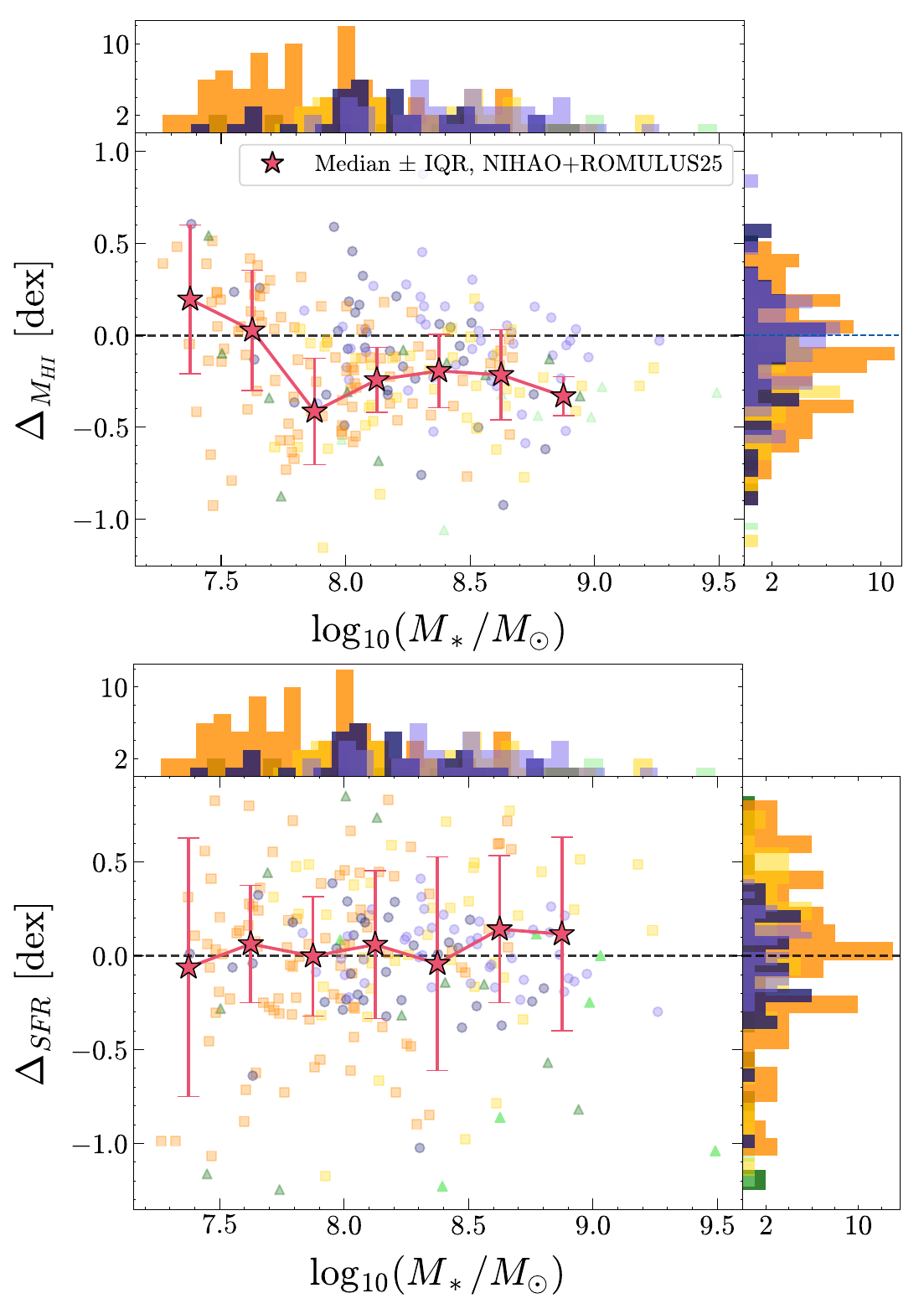}
    \caption{Residuals $\Delta \, M_{HI}$ between the comparison samples and the \citet{2015ApJ...809..146B} relation plotted in Figure \ref{fig:mhimstarsfr} (top), and residuals $\Delta \, SFR$ with respect to the best-fitting line for SMUDGes-HI UDGs and dwarfs combined (bottom).  The symbols are the same as in Figure \ref{fig:mhimstarsfr}. The red stars and bars show the medians and IQRs of the residuals for the \textsc{Romulus25} and \textsc{NIHAO}  samples combined, and histograms on both axes show the distributions of the samples. The median simulated \textsc{Romulus25} and \textsc{NIHAO} systems have lower $M_{HI}$ and a larger scatter in  in $SFR$ than in the SMUDGes observations.}
    \label{fig:resplotcomb}
\end{figure*}

While there is also strong overlap between the SMUDGes-HI dwarfs and UDGs and the trend measured by \citet{2015ApJ...809..146B} for optically-selected field dwarfs, the locus of the \textsc{NIHAO} and \textsc{Romulus25} samples falls below it. This is highlighted in the top panel in Figure \ref{fig:resplotcomb}, which shows the residuals in $M_{HI}$ of the points in Figure \ref{fig:mhimstarsfr} relative to the \citet{2015ApJ...809..146B} relation. In Figure \ref{fig:resplotcomb}, the red stars and error bars show the medians and interquartile ranges (IQR) of the residuals for \textsc{Romulus25} and \textsc{NIHAO} combined. For $\log({M_*/M_\odot}) \gtrsim 8$, $M_{HI}$ for the simulated galaxies lies $\sim 0.25\,$dex (i.e.\ a factor of two) below the observed trends, and a PQMass test comparing SMUDGes-HI dwarfs+UDGs to \textsc{Romulus25} dwarfs+UDGs confirms that, statistically, they are not drawn from the same distribution ($p \sim 10^{-20}$). We discuss this further in Section~\ref{sec:q3}.

\subsubsection{$SFR-M_*$ Relation}
In the bottom panel of Figure \ref{fig:mhimstarsfr}, we compare the distribution of simulated and observed galaxies in $SFR$ and $M_*$. The SMUDGes-HI dwarfs and UDGs follow a tight relation in this space, and are consistent with being drawn from the same distribution (PQMass $p \sim 0.06$). They sit below the best fitting star forming main sequence (SFMS) for massive, optically-selected field galaxies from \citet{saintonge22}, although the overlap in $M_*$ between that sample and ours is low. Since our comparison samples are selected to have low surface brightnesses, it is not surprising that they sit below the SFMS defined by higher surface brightness objects \citep{2011AdAst2011E..12S}. 

 While the simulated galaxies follow a similar $SFR - M_*$ relation as the observations, they display significantly higher scatter in $SFR$. This is highlighted in the bottom panel of Figure \ref{fig:resplotcomb}, which shows the residuals in $SFR$ of the points in Figure \ref{fig:mhimstarsfr} with respect to the line-of-best-fit to the SMUDGes-HI dwarfs and UDGs combined (c.f.\ Figure~\ref{fig:mhimstarsfr}). The median residuals of the \textsc{nihao} +\textsc{Romulus25} galaxies in the bottom panel of Figure~\ref{fig:resplotcomb} (red stars) follow the zero line closely, showing that the simulations are consistent with the observed trend. However, the IQR of the simulations (red error bars) is $\sim 1$ dex for most bins, significantly exceeding the scatter in the SMUDGes-HI dwarfs+UDGs (PQMass $p \sim 10^{-28}$).   We note that this difference between the simulations and observations under-estimates that in intrinsic scatter between these samples, because observational errors contribute to the scatter in the SMUDGes points. We therefore find that the simulated galaxies in our comparison samples have much more variable SFRs at a given stellar mass than the observations; we discuss this result further in Section~\ref{sec:q2}. 

 In contrast to the similarity between the distributions of \textsc{Romulus25} dwarfs and UDGs in $M_{HI} - M_*$ (Figure~\ref{fig:mhimstarsfr} top), a PQMass test applied to these two simulated comparison samples indicates that they are significantly different in $SFR - M_*$ space ($p \sim 10^{-5}$). Examining the $M_*$ histograms in Figure~\ref{fig:resplotcomb}, we speculate that this difference arises from the excess \textsc{Romulus25} UDGs relative to \textsc{Romulus25} dwarfs at low $M_*$, and vice-versa at high $M_*$. A PQMass test indicates that the \textsc{Romulus25} dwarfs and UDGs are consistent with being drawn from the same distribution for $8.0 \leq \log(M_*/M_\odot) \leq 8.8$ ($p \sim 0.16$), supporting this hypothesis. We note that this difference in $M_*$ distribution is present to a lesser extent in the observed dwarfs and UDGs, as well; we discuss this result further in Section~\ref{sec:q1}.

\begin{figure}
    \centering
    \includegraphics[width=0.45\textwidth]{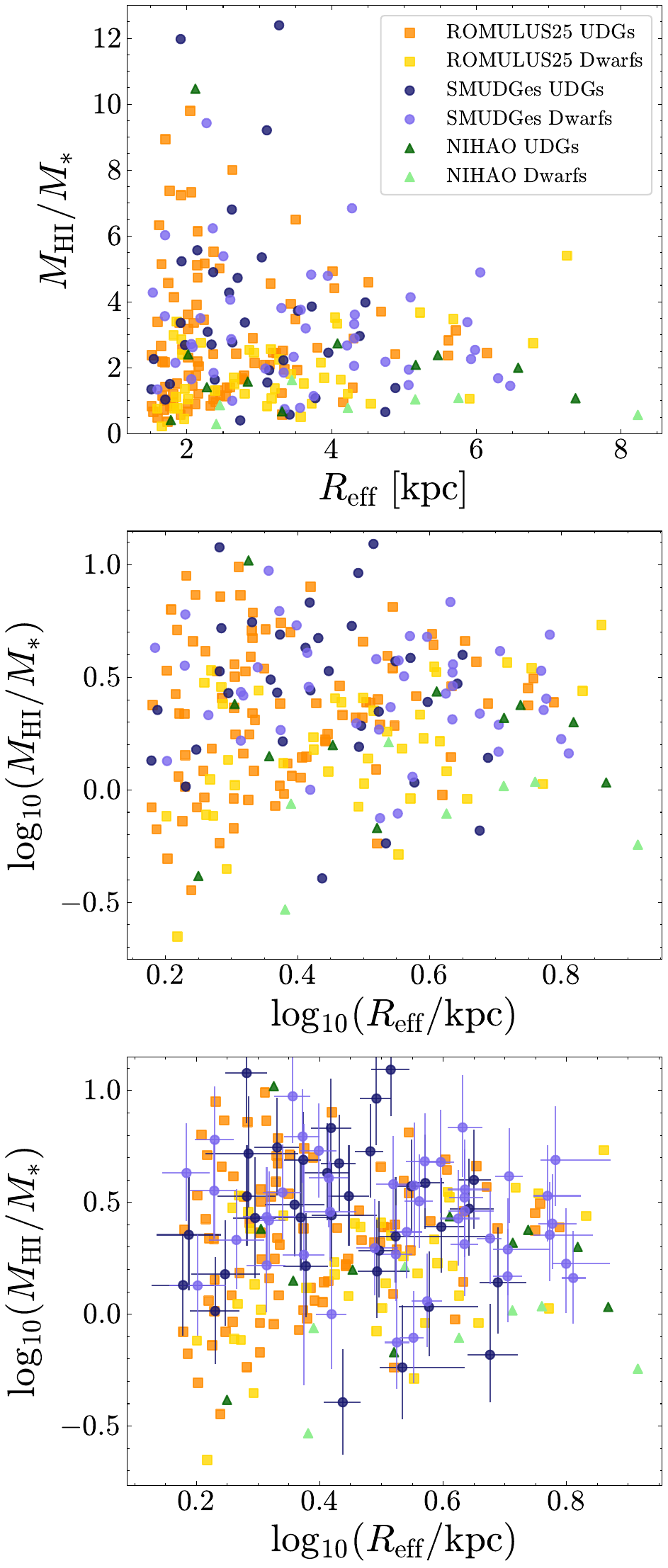}
    \caption{Distribution of gas fraction ($M_{\mathrm{HI}}/M_*$) as a function of the effective radius $R_{\mathrm{eff}}$ of UDGs and dwarfs in linear space (top), in log space (middle), and in log space with uncertainties on the SMUDGes data points overplotted (bottom), using the same symbol shapes and colours as in Figures~\ref{fig:mhimstarsfr}~and~\ref{fig:resplotcomb}. Regardless of the representation adopted, the distributions of observed and simulated UDGs and dwarfs are well-mixed in the $M_{\mathrm{HI}}/M_*$ -- $R_{\mathrm{eff}}$ plane.}
    \label{fig:grsize}
\end{figure}

\subsection{Gas Richness and Size}\label{sec:res-gr}

We now examine the relationship between gas richness and size for our comparison samples. This parameter space is of particular interest given the findings from \textsc{nihao} by \citet{di2017nihao} that, at a given stellar mass, UDGs with higher gas richnesses have larger sizes in comparison to dwarfs. 

Figure \ref{fig:grsize} shows the distributions of gas richness $M_{\mathrm{HI}}/M_*$ as a function of optical half-light radius $R_{\mathrm{eff}}$ for the simulated and observed comparison samples using the same symbol types and colors as in Figures~\ref{fig:mhimstarsfr}~and~\ref{fig:resplotcomb} (we show and discuss median trends in Figure \ref{fig:grsizebinsmed} below). To highlight different trends, we plot these distributions in linear space (top) and in log space (middle and bottom), with the bottom panel including uncertainties on the observations. In general,  there is little qualitative difference between UDGs and dwarfs in either the simulated or observed samples -- they are strongly intermixed. 

We note that the overabundance of low-$M_*$ UDGs in \textsc{Romulus25} evident in the upper histograms in Figure~\ref{fig:resplotcomb} appears as an excess of low-$R_{eff}$ UDGs in Figure \ref{fig:grsize}; this is most obvious in linear space (top panel). While many of these UDGs have $M_{\mathrm{HI}}/M_* \lesssim 2$ and overlap with \textsc{Romulus25} dwarfs, there is an excess of extremely gas-rich ($M_\mathrm{HI}/M_* > 4$) UDGs which have no simulated dwarf counterparts. In Section \ref{sec:q1}, we discuss why we think this may be. 

There are hints of a positive trend between $M_{\mathrm{HI}}/M_*$ and $R_{eff}$ in Figure \ref{fig:grsize}, which is most evident in log space (middle and bottom panels). 
We explore this in more detail in Figure \ref{fig:grsizebinsmed}, dividing the total (left), observed (middle) and simulated (right) comparison samples into equally-sized bins in $M_*$ to control for the well-known correlation between gas richness and stellar mass \citep[e.g.][]{saintonge22}. We also plot the medians and IQRs of all galaxies within each $M_*$ bin (stars with error bars connected by solid lines), where the ranges are chosen to produce medians over equal numbers of galaxies.  We also plot medians and IQRs for UDGs and dwarfs in each stellar mass bin separately (dashed and dotted lines, respectively), provided there are at least four such points in that range. 

Figure \ref{fig:grsizebinsmed} clearly shows that, as expected, galaxies with lower $M_*$ are more gas-rich; this is confirmed by PQMass tests ($p < 0.001$ between the mass bins in each panel). There is also a clear positive correlation between gas-richness and size, with objects that are physically larger within a given mass bin being more gas-rich, for UDGs and dwarfs in both simulations and observations.  The trend is less obvious in the observations alone (middle panel) than in the simulations alone (right panel) \footnote{We consider \textsc{Romulus25} and \textsc{nihao} separately in Figure \ref{fig:linmix}.}. Although there are some differences between the medians and IQRs for dwarfs (dotted lines) and UDGs (dashed) within a given mass bin, they are relatively small; indeed, PQMtests show that UDGs and dwarfs are not distinct in either observations or simulations, or even when they're considered collectively as in the left panel of Figure \ref{fig:grsizebinsmed}.

    

 In order to quantify the trends traced by the medians in Figure \ref{fig:grsizebinsmed}, we obtain best-fit linear relations in $\log M_{HI}-\log R_{eff}$ space using the the Python implementation of \verb|LinMix| \citep{2007ApJ...665.1489K}, as done in \citet{karunakaran2024}. Briefly, \verb|LinMix| is a Bayesian fitting routine which derives a likelihood function for the data.
  Figure \ref{fig:linmix} shows the resulting slopes and intercepts of the best fit lines to the observed and simulated samples shown in Figure \ref{fig:grsizebinsmed} (left and right panels), as well as to the \textsc{Romulus25} sample alone (middle panel). The best-fit lines to observed galaxies in each stellar mass bin are consistent within scatter. There is also no statistical difference between observed UDGs and dwarfs in each stellar mass bin. When looking at \textsc{Romulus25} alone, the slopes of the different mass bins are consistent and lie within a range of around 1.0 and 1.75 (not including scatter). With the addition of \textsc{nihao}, the range in slopes increases to between 0 and 2. Overall, we don't see a statistical difference in the slopes of UDGs and dwarfs in either simulations nor observations.



\begin{figure*}[hbt!]
    \centering
    \includegraphics[width=\textwidth]{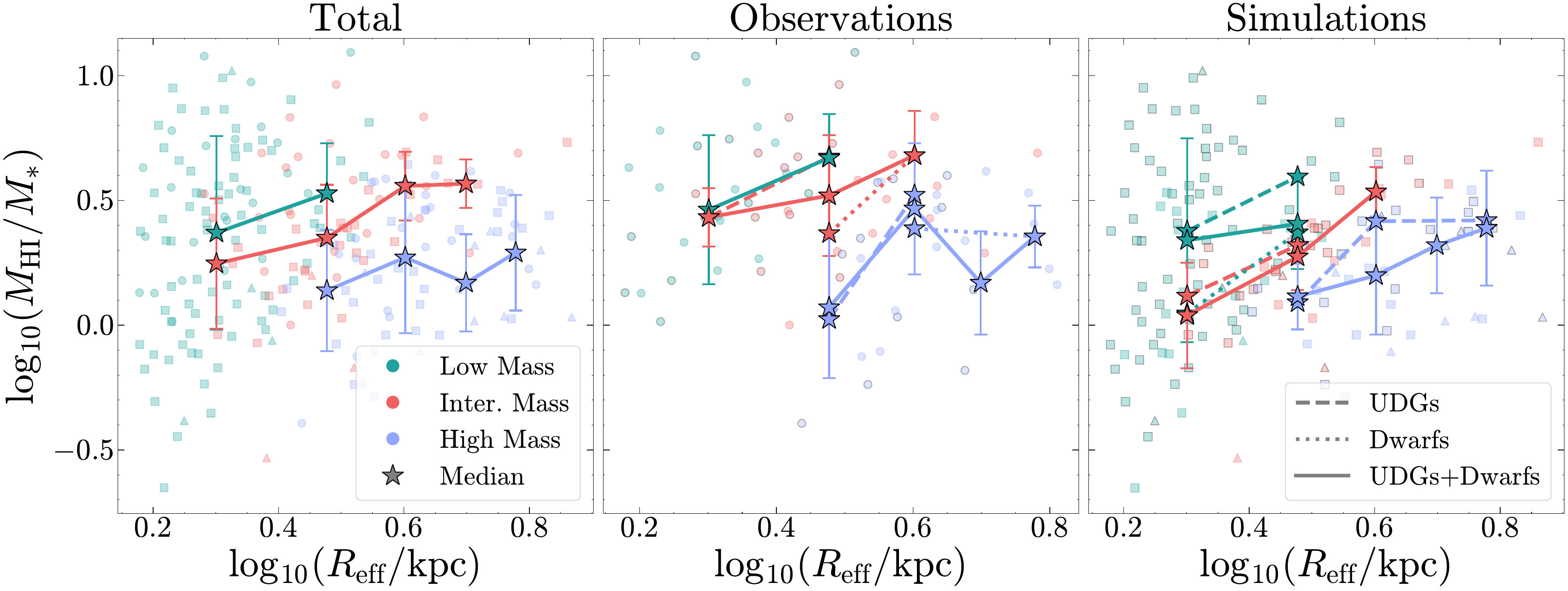}
    \caption{Gas richness ($M_{HI}/M_*$) as a function of effective radius ($R_{eff}$) for all UDGs and dwarfs in the comparison samples (left), and then divided into observed SMUDGes-HI UDGs and dwarfs (middle) and simulated \textsc{nihao} and \textsc{Romulus25} UDGs and dwarfs (right). In all panels, the symbol types are the same as in Figures~\ref{fig:mhimstarsfr}~--~\ref{fig:grsize}, but the colours divide UDGs and dwarfs into similar-sized $M_*$ bins (teal, red and purple points show $\log_{10}(M_*/M_\odot)$ bins from $7.3 - 8.0$, $8.0 - 8.2$, $8.2 - 8.9$ for UDGs, and $7.4 - 8.3$, $8.3 - 8.6$, $8.6 - 9.3$ for dwarfs). Stars show the medians and IQRs within each $M_*$ bin, which are connected by solid lines for UDGs and dwarfs combined, dashed lines for only UDGs, and dotted lines for dwarfs only. $M_{HI}/M_*$ is correlated with $R_{eff}$ for both dwarfs and UDGs in both simulations and observations.}
    \label{fig:grsizebinsmed}
\end{figure*}


\section{Discussion}\label{sec:discussion}
In this section, we return to our original question: are optically-selected gas-rich UDGs and dwarfs in our sample distinct in simulations or observations? We also consider two complementary questions: how well do simulated UDGs agree with observed UDGs? And how do the properties of UDGs and dwarfs formed in the \textsc{Romulus25} simulations compare with those formed in \textsc{nihao}? Below, we consider each question in turn.

\subsection{Comparing UDGs and Dwarfs in Simulations and Observations}\label{sec:q1}

It has been established in both observations (e.g. \citealt{karunakaran2024}) and simulations (e.g. \citealt{2022ApJ...926...92V}) that the definitions used to separate UDGs from dwarf galaxies have significant effects on subsequent analyses of their properties. We first address how these definitions can affect our interpretations.

While many authors require that UDGs have $R_{\rm{eff}} > 1.5$ kpc, others (e.g. \citealt{di2017nihao, cardona2020nihao}) adopt the more relaxed definition of $R_{\rm{eff}} > 1$ kpc. As summarized in Table \ref{tab:cuts}, a significant number of galaxies in our \textsc{nihao} sample were cut from the parent sample due to our more conservative UDG definition of $R_{\rm{eff}} > 1.5$ kpc (in addition to the requirement of being actively star-forming). As a result, the UDGs in our \textsc{nihao} sample are significantly fewer than those in the works of \citet{di2017nihao} or \citet{cardona2020nihao}. As a check, we repeat our analyses with the relaxed definition of $R_{\rm{eff}} > 1$ kpc and find that it does not significantly affect our overall results. We also repeat our analyses to include dwarfs that are on the edge of the surface-brightness criterion as UDGs and again, find no difference in our results. This further highlights how UDGs and dwarfs have a continuum of properties rather than being two bimodal populations. 

Moreover, the simulations have more small UDGs than large ones and more small UDGs than small dwarfs. This could be in large part due to resolution effects: \textsc{Romulus25} is known to over-quench dwarfs below $M_* \sim 10^8~\mathrm{M}_\odot$ resulting in them being removed from our final sample \citep{wright2021formation}. UDGs and dwarfs in our observed and simulated samples follow the same $M_{HI}-M_*$ relation. Our line fits to the UDGs and dwarfs in each of the three samples are consistent within uncertainty, i.e. there is no statistical difference in UDG and dwarfs in $M_{HI}$ in neither observations nor simulations. 

\begin{figure*}
    \centering
    \includegraphics[width=\linewidth]{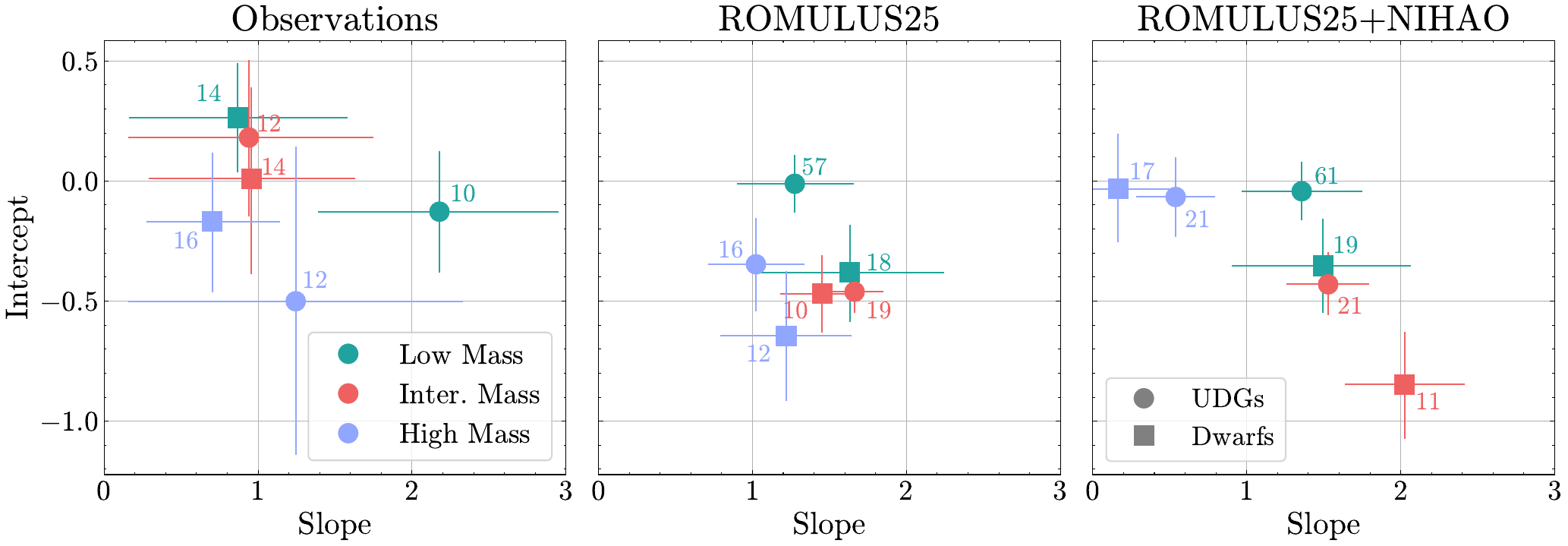}
    \caption{Slopes and intercepts of best-fit lines from the LinMix fitting procedure to the $M_\mathrm{HI}/M_* - R_{eff}$ relations shown in Figure~\ref{fig:grsizebinsmed} for SMUDGes observations (left), the \textsc{Romulus25} comparison sample (middle), and both \textsc{Romulus25} and \textsc{nihao} collectively (right).
     In all panels, the number beside each UDG (circle) and dwarf (square) symbol is the number of galaxies present in that mass bin. With the possible exception of the low-mass and high-mass dwarfs in \textsc{Romulus25} and \textsc{nihao} combined, there is no significant difference in $M_\mathrm{HI}/M_* - R_{eff}$ for dwarfs and UDGs in the simulations or in the observations.}
    \label{fig:linmix}
\end{figure*}

However, UDGs and dwarfs do not span the same extents in stellar mass. In both observations and simulations, we notice that there are no dwarfs present below $M_* \lesssim 10^{7.7}~\rm{M}_\odot$. Additionally, the UDGs present in this mass range are extremely gas-rich. \citet{wright2021formation} note that the lack of low-mass dwarf galaxies in the simulations where observations exist (e.g. \citealt{2002ASPC..273..341S, zitrin2009star}) is a known problem, not just in \textsc{Romulus25}, but in many cosmological simulations, including those with considerably higher resolution (e.g. \citealt{santos2018nihao,garrison2019local}). Close to the simulations' resolution limit ($M_*\sim 10^{7.5}~\rm{M}_\odot$), small structures are not resolved, influencing galaxy sizes and contributing to the formation of UDGs - and not dwarfs - at these lower masses. As well, \textsc{Romulus25} over-quenches its dwarfs below $M_* \sim 10^8~\rm{M}_\odot$, in part due to the resolution constraint of initial supermassive black hole seeds being fairly high mass. The same processes that result in complete quenching will also tend to reduce the gas content and SFR of non-quenched galaxies in that mass range. As a result, these galaxies are probably redder (and therefore optically fainter) than we might expect. Thus, we decide not to over-interpret the overabundance of UDGs in \textsc{Romulus25} where no dwarfs exist. We see the same effect in \textsc{nihao} but here, it is unlikely to be a resolution effect given that \textsc{nihao} is a zoom-in simulation. However, because it is a zoom-in simulation, it does not cover a volume large enough to be representative of the population of UDGs and dwarfs. The low statistics mean that we should be wary of making population-level inferences. Similarly, the lack of low-mass dwarfs in SMUDGes may be due to low statistics in that regime. Given all these caveats, we refrain from over-interpreting this result. 

Between $10^{7.7} \lesssim M_*/\rm{M}_\odot \lesssim 10^{9}$ we see roughly equal UDGs and dwarfs in simulations and observations. However, above $M_* \gtrsim 10^{9}~\rm{M}_\odot$, we see no UDGs in observations and simulations where we see dwarfs. This is expected; more mass within a given area of the galaxy will result in an increase in the central surface brightness, pushing the galaxy out of the UDG regime. 

We see a positive correlation between gas-richness and size. This is a feature of the sample and different from the typical trend seen over a large range in mass and samples not selected on surface brightness \citep{2014ApJ...793...40H}. We also see a positive correlation of size with stellar mass, i.e. larger disks tend to be more massive and have a higher gas fraction. Statistically, we find no difference in the gas-richness vs size relation that UDGs follow compared to dwarfs, in either simulations or observations. However, we may pick out by eye in Figure \ref{fig:grsizebinsmed} that UDGs in simulations are slightly more gas-rich than dwarfs, although the same effect is not seen in observations. This is interesting, given that simulations like \textsc{nihao} \citep{di2017nihao} specifically predict that for a given stellar mass, the larger the effective radius the larger the gas reservoir, and therefore, UDGs generally have higher gas-fractions when compared to dwarfs. 

\subsection{Comparing UDGs in Observations and Simulations}\label{sec:q2}
UDGs and dwarfs in simulations cover the same regions in parameter space as observations. Figure \ref{fig:sbreffful} shows that all three samples overlap not only in the extent of our comparison samples but also in the parent samples. However, we see one difference in the observed and simulated samples: the simulations predict the presence of highly extended ($R_{\rm{eff}} > 5$ kpc) UDGs, while we do not observe any in SMUDGes. While this may be a result of low statistics, we also speculate that it may be because the largest (and, therefore, most massive) UDGs are rare \citep{van2017abundance}. While UDGs are certainly outliers to the mass-size relation (e.g. \citealt{ van2013assembly,lange2015galaxy}), it appears that both \textsc{nihao} and \textsc{Romulus25} produce field UDGs that are more diffuse than observed.

While UDGs and dwarfs have consistent $M_{HI}$ in each simulated and observed sample, we see a difference between the gas-richness of UDGs and dwarfs in simulations versus in observations. Many properties of our SMUDGes sample are largely consistent with other UDG samples (see \citealt{karunakaran2024}) e.g. HI-selected samples like the ALFALFA HUDs \citep{leisman2017almost, kado2022ultra} but they are systematically less gas-rich than the HUDs. This is expected as HI-selected surveys tend to be biased high compared to optically selected surveys (e.g. \citealt{catinella2010galex}). \textsc{nihao} UDGs are less gas-rich than SMUDGes and the HUDs. Figure \ref{fig:resplotcomb} demonstrates how the the gas-richness of simulated galaxies in \textsc{Romulus25} and \textsc{nihao} lies systematically below that of SMUDGes. \citet{wright2021formation} see the same result in their sample of \textsc{Romulus25} UDGs and believe the discrepancy is caused by differences in how $R_{\rm{eff}}$ is calculated in simulations versus observations. In simulations, galaxies are rotated to be face-on before fitting Sérsic profiles, resulting in the maximum $R_{\rm{eff}}$ possible. While the SMUDGes observations are also fit with the assumption that the galaxies are face-on, the inclinations are poorly constrained, making our $R_{\rm{eff}}$ values likely underestimates. Adjusting these values higher may align the samples better. Additionally, we speculate that in \textsc{nihao}, the periodic bursts of intense star formation may drive large amounts of gas out of the galaxy, resulting in systematically lower gas-richness.

The effects of bursty star formation may also influence the $SFR-M_*$ relation in the bottom panels of Figure \ref{fig:mhimstarsfr} and Figure \ref{fig:resplotcomb}. While UDGs and dwarfs in simulations and observations follow the same relation, the simulations have much more variable SFR at any given value of $M_*$, even when accounting for the SMUDGes uncertainties. This could be the result of an ``over-bursty" prescription for the stochastic star formation that builds stellar mass in dwarf galaxies: while the stellar mass matches observations in the mean, star formation is more rigorous at its peak and more quiescent at its minimum than observed. 
 These findings are consistent with the results from \citet{2021MNRAS.500.1414B} and \citet{arora2023manga}, who show that low mass galaxies in \textsc{nihao} have larger scatter in the $SFR - M_*$ plane than high mass galaxies. 

\subsection{Comparison of UDGs in \textsc{nihao} and \textsc{Romulus25}}\label{sec:q3}
The broad correspondence between the observed and simulated UDGs in our study demonstrates that these systems form naturally in both \textsc{nihao} and \textsc{Romulus25}, without requiring exotic physics (e.g., \citealt{2024A&A...690A.149N, 2024arXiv240311403Z}). However, they may form through different mechanisms: in \textsc{nihao}, UDG progenitors expand as a result of episodic star formation at early times \citep{dicintio2019}, while in \textsc{Romulus25}, they are the product of major mergers \citep{wright2021formation}. However, these two processes may be happening concurrently in the  simulations. The resolution of \textsc{Romulus25} is not sufficient to achieve the densities necessary for star formation to produce dark matter cores \citep{wright2021formation}, so if it were contributing to the formation of UDGs in \textsc{Romulus25} we would not have the information. Similarly, the small number of \textsc{nihao} zoom-in simulations, may imply that the effect of mergers is not fully captured. On the other hand, since \textsc{nihao} and \textsc{Romulus25} use similar hydro-solvers, it is perhaps not surprising that the UDGs they form display such similar properties.  

Many authors (e.g. \citealt{ferre2018origins}) have suggested that the variety of observations and theoretical predictions that support competing formation models may imply that UDGs can form through several different mechanisms.
Thus, it is very interesting that UDGs from both simulations examined here have such consistent properties, such as slopes in gas richness-size space (see Figure \ref{fig:linmix}). We find no significant difference in the two samples in any of the parameter spaces that we investigated. 
Our results suggest that, if there is more than one mechanism at work to produce gas-rich field UDGs, then their effects are indistinguishable in the parameter distributions that we explore here and indistinguishable from those of dwarfs. 

It is possible that there exists complexity of structural and evolutionary differences between UDGs and dwarfs that is hidden from the properties that we have examined (see \citealt{2021ApJ...920...72K,2022ApJ...926...92V}). However, the simplest explanation of our results is that gas-rich dwarfs and UDGs have the same continuum of properties, and form from the same combination of mechanisms, rather than being distinct populations.


\section{Summary and Conclusions}\label{sec:conclusion}
In this paper, we compare observed UDGs and dwarfs from the SMUDGes-HI survey with simulated UDGs and dwarfs in the \textsc{nihao} and \textsc{Romulus25} simulations. From these parent samples, we select only star-forming ($sSFR \geq 10^{-11.5}~\mathrm{yr}^{-1}$), gas-rich ($f_{\rm HI} > 0.2$), and extended ($1.5 < R_{\rm{eff}}/\rm{kpc} < 10$) galaxies that would match the surface brightnesses ($\mu_{\mathrm{0,g}} > 23.5$ mag arcsec$^{-2}$) of the galaxies we observe in SMUDGes. We compare UDGs and dwarfs in several parameter spaces, including $\mu_{0,g}-R_{\rm{eff}}$, $M_{HI}-M_*$, $SFR-M_*$, and $M_{HI}/M_*-R_{\rm{eff}}$. We summarize the results of our comparisons below:

\begin{itemize}
    \item We find no statistically significant difference in UDGs and dwarfs in neither observations nor simulations. However, we see hints in $M_{HI}/M_*-R_{\rm{eff}}$ space that UDGs in simulations may be slightly more gas-rich than dwarfs. UDGs and dwarfs displaying similar properties in observations and simulations may indicate that they are likely not distinct populations. The simplest explanation seems to be that they form through similar mechanisms, with UDGs undergoing a more extreme version of those mechanisms than dwarfs. 
    \item UDGs and dwarfs in simulations are remarkably consistent with observations. However, we find that simulated galaxies are systematically less gas-rich than observations by about ${\sim}0.25$ dex, which, we speculate might be related to the vigorous stellar feedback employed in both the simulations analyzed in this study. Moreover, simulated galaxies display much larger scatter (${\sim}1$ dex) in SFR than observations. This might point to some general limitations on the ability to capture the complex network of star formation and feedback in cosmological simulations \citep{vogelsberger2020cosmological}.
    \item As predicted by the \textsc{nihao} simulations \citep{di2017nihao}, we see a positive correlation between gas-richness and size in both observations and simulations. However, unlike what is predicted by \textsc{nihao}, we do not see a statistical difference in gas-richness in UDGs and dwarfs, in neither simulations nor observations. In addition, the same trends in gas richness and size are produced by the \text{Romulus25} simulations, which form UDGs through early major mergers \citep{wright2021formation}. 
    \item UDGs in the \textsc{nihao} and \text{Romulus25} simulations display remarkably consistent properties, despite forming them through different mechanisms. If one or both of these mechanisms are occurring in nature, then their effects are indistinguishable, given our current observables. 
\end{itemize}

From the perspective of the HI masses, gas richnesss, star formation rates and optical sizes that we have examined in this paper, our results suggest that gas-rich, star forming UDGs and dwarfs are not distinct galaxy populations, either observationally or in simulations. This adds to the growing body of work, on other observables for other UDG populations, which reach similar conclusions \citep{2016ApJ...819L..20B,Conselice_2018,2023MNRAS.519.1545C,2024ApJS..271...52Z,Buzzo25}. UDGs and dwarfs may be distinct in some other observable, or at another cosmic time. One avenue for exploring this possibility is to examine the evolution of simulated systems that are classified as dwarfs and UDGs at $z=0$; this work is now underway.  



\begin{acknowledgments}

This work makes use of data obtained with the Robert C. Byrd Green Bank Telescope, supported by the National Radio Astronomy Observatory and Green Bank Observatory which are major facilities funded by the U.S. National Science Foundation operated by Associated Universities, Inc.
The NIHAO simulations are based upon work supported by Tamkeen under the NYU Abu Dhabi Research Institute grant CASS.
The authors gratefully acknowledge the Gauss Centre for Supercomputing e.V. (www.gauss-centre.eu) for enabling this project via computing time on the GCS Supercomputer SuperMUC at Leibniz Supercomputing Centre (www.lrz.de) along with the high  performance computing resources at New York University Abu Dhabi.

This research was undertaken thanks in part to funding from the Canada First Research Excellence Fund through the Arthur B. McDonald Canadian Astroparticle Physics Research Institute.
KS acknowledges funding from the Natural Sciences and Engineering Research Council of Canada (NSERC).

\end{acknowledgments}



\newpage
\bibliography{bib}{}

\begin{thebibliography}{}
\expandafter\ifx\csname natexlab\endcsname\relax\def\natexlab#1{#1}\fi
\providecommand{\url}[1]{\href{#1}{#1}}
\providecommand{\dodoi}[1]{doi:~\href{http://doi.org/#1}{\nolinkurl{#1}}}
\providecommand{\doeprint}[1]{\href{http://ascl.net/#1}{\nolinkurl{http://ascl.net/#1}}}
\providecommand{\doarXiv}[1]{\href{https://arxiv.org/abs/#1}{\nolinkurl{https://arxiv.org/abs/#1}}}

\bibitem[{{Abel} {et~al.}(1997){Abel}, {Anninos}, {Zhang}, \& {Norman}}]{1997NewA....2..181A}
{Abel}, T., {Anninos}, P., {Zhang}, Y., \& {Norman}, M.~L. 1997, \na, 2, 181, \dodoi{10.1016/S1384-1076(97)00010-9}

\bibitem[{Abraham \& van Dokkum(2014)}]{abraham2014ultra}
Abraham, R.~G., \& van Dokkum, P.~G. 2014, Publications of the Astronomical Society of the Pacific, 126, 55

\bibitem[{{Aihara} {et~al.}(2018){Aihara}, {Arimoto}, {Armstrong}, {Arnouts}, {Bahcall}, {Bickerton}, {Bosch}, {Bundy}, {Capak}, {Chan}, {Chiba}, {Coupon}, {Egami}, {Enoki}, {Finet}, {Fujimori}, {Fujimoto}, {Furusawa}, {Furusawa}, {Goto}, {Goulding}, {Greco}, {Greene}, {Gunn}, {Hamana}, {Harikane}, {Hashimoto}, {Hattori}, {Hayashi}, {Hayashi}, {He{\l}miniak}, {Higuchi}, {Hikage}, {Ho}, {Hsieh}, {Huang}, {Huang}, {Ikeda}, {Imanishi}, {Inoue}, {Iwasawa}, {Iwata}, {Jaelani}, {Jian}, {Kamata}, {Karoji}, {Kashikawa}, {Katayama}, {Kawanomoto}, {Kayo}, {Koda}, {Koike}, {Kojima}, {Komiyama}, {Konno}, {Koshida}, {Koyama}, {Kusakabe}, {Leauthaud}, {Lee}, {Lin}, {Lin}, {Lupton}, {Mandelbaum}, {Matsuoka}, {Medezinski}, {Mineo}, {Miyama}, {Miyatake}, {Miyazaki}, {Momose}, {More}, {More}, {Moritani}, {Moriya}, {Morokuma}, {Mukae}, {Murata}, {Murayama}, {Nagao}, {Nakata}, {Niida}, {Niikura}, {Nishizawa}, {Obuchi}, {Oguri}, {Oishi}, {Okabe}, {Okamoto}, {Okura}, {Ono}, {Onodera}, {Onoue}, {Osato}, {Ouchi}, {Price}, {Pyo},
  {Sako}, {Sawicki}, {Shibuya}, {Shimasaku}, {Shimono}, {Shirasaki}, {Silverman}, {Simet}, {Speagle}, {Spergel}, {Strauss}, {Sugahara}, {Sugiyama}, {Suto}, {Suyu}, {Suzuki}, {Tait}, {Takada}, {Takata}, {Tamura}, {Tanaka}, {Tanaka}, {Tanaka}, {Tanaka}, {Terai}, {Terashima}, {Toba}, {Tominaga}, {Toshikawa}, {Turner}, {Uchida}, {Uchiyama}, {Umetsu}, {Uraguchi}, {Urata}, {Usuda}, {Utsumi}, {Wang}, {Wang}, {Wong}, {Yabe}, {Yamada}, {Yamanoi}, {Yasuda}, {Yeh}, {Yonehara}, \& {Yuma}}]{2018PASJ...70S...4A}
{Aihara}, H., {Arimoto}, N., {Armstrong}, R., {et~al.} 2018, \pasj, 70, S4, \dodoi{10.1093/pasj/psx066}

\bibitem[{{Amorisco} \& {Loeb}(2016)}]{2016MNRAS.459L..51A}
{Amorisco}, N.~C., \& {Loeb}, A. 2016, \mnras, 459, L51, \dodoi{10.1093/mnrasl/slw055}

\bibitem[{Arora {et~al.}(2023)Arora, Courteau, Stone, \& Macci{\`o}}]{arora2023manga}
Arora, N., Courteau, S., Stone, C., \& Macci{\`o}, A.~V. 2023, Monthly Notices of the Royal Astronomical Society, 522, 1208

\bibitem[{{Arora} {et~al.}(2022){Arora}, {Macci{\`o}}, {Courteau}, {Buck}, {Libeskind}, {Sorce}, {Brook}, {Hoffman}, {Yepes}, {Carlesi}, \& {Stone}}]{Arora2022}
{Arora}, N., {Macci{\`o}}, A.~V., {Courteau}, S., {et~al.} 2022, \mnras, 512, 6134, \dodoi{10.1093/mnras/stac893}

\bibitem[{{Beasley} {et~al.}(2016){Beasley}, {Romanowsky}, {Pota}, {Navarro}, {Martinez Delgado}, {Neyer}, \& {Deich}}]{2016ApJ...819L..20B}
{Beasley}, M.~A., {Romanowsky}, A.~J., {Pota}, V., {et~al.} 2016, \apjl, 819, L20, \dodoi{10.3847/2041-8205/819/2/L20}

\bibitem[{Bellazzini {et~al.}(2017)Bellazzini, Belokurov, Magrini, Fraternali, Testa, Beccari, Marchetti, \& Carini}]{bellazzini2017redshift}
Bellazzini, M., Belokurov, V., Magrini, L., {et~al.} 2017, Monthly Notices of the Royal Astronomical Society, 467, 3751

\bibitem[{{Bennet} {et~al.}(2017){Bennet}, {Sand}, {Crnojevi{\'c}}, {Spekkens}, {Zaritsky}, \& {Karunakaran}}]{2017ApJ...850..109B}
{Bennet}, P., {Sand}, D.~J., {Crnojevi{\'c}}, D., {et~al.} 2017, \apj, 850, 109, \dodoi{10.3847/1538-4357/aa9180}

\bibitem[{{Blank} {et~al.}(2019){Blank}, {Macci{\`o}}, {Dutton}, \& {Obreja}}]{Blank2019}
{Blank}, M., {Macci{\`o}}, A.~V., {Dutton}, A.~A., \& {Obreja}, A. 2019, \mnras, 487, 5476, \dodoi{10.1093/mnras/stz1688}

\bibitem[{Blank {et~al.}(2019)Blank, Macci{\`o}, Dutton, \& Obreja}]{blank2019nihao}
Blank, M., Macci{\`o}, A.~V., Dutton, A.~A., \& Obreja, A. 2019, Monthly Notices of the Royal Astronomical Society, 487, 5476

\bibitem[{{Blank} {et~al.}(2021{\natexlab{a}}){Blank}, {Meier}, {Macci{\`o}}, {Dutton}, {Dixon}, {Soliman}, \& {Kang}}]{Blank2021}
{Blank}, M., {Meier}, L.~E., {Macci{\`o}}, A.~V., {et~al.} 2021{\natexlab{a}}, \mnras, 500, 1414, \dodoi{10.1093/mnras/staa2670}

\bibitem[{{Blank} {et~al.}(2021{\natexlab{b}}){Blank}, {Meier}, {Macci{\`o}}, {Dutton}, {Dixon}, {Soliman}, \& {Kang}}]{2021MNRAS.500.1414B}
---. 2021{\natexlab{b}}, \mnras, 500, 1414, \dodoi{10.1093/mnras/staa2670}

\bibitem[{Bothun {et~al.}(1991)Bothun, Impey, \& Malin}]{bothun1991extremely}
Bothun, G.~D., Impey, C.~D., \& Malin, D.~F. 1991, Astrophysical Journal, Part 1 (ISSN 0004-637X), vol. 376, Aug. 1, 1991, p. 404-423., 376, 404

\bibitem[{{Bradford} {et~al.}(2015){Bradford}, {Geha}, \& {Blanton}}]{2015ApJ...809..146B}
{Bradford}, J.~D., {Geha}, M.~C., \& {Blanton}, M.~R. 2015, \apj, 809, 146, \dodoi{10.1088/0004-637X/809/2/146}

\bibitem[{{Bromm} {et~al.}(2001){Bromm}, {Kudritzki}, \& {Loeb}}]{2001ApJ...552..464B}
{Bromm}, V., {Kudritzki}, R.~P., \& {Loeb}, A. 2001, \apj, 552, 464, \dodoi{10.1086/320549}

\bibitem[{{Buzzo} {et~al.}(2025){Buzzo}, {Forbes}, {Jarrett}, {Marleau}, {Duc}, {Brodie}, {Romanowsky}, {Ferr{\'e}-Mateu}, {Hilker}, {Gannon}, {Pfeffer}, \& {Haacke}}]{Buzzo25}
{Buzzo}, M.~L., {Forbes}, D.~A., {Jarrett}, T.~H., {et~al.} 2025, \mnras, 536, 2536, \dodoi{10.1093/mnras/stae2700}

\bibitem[{{Cardona-Barrero} {et~al.}(2023){Cardona-Barrero}, {Di Cintio}, {Battaglia}, {Macci{\`o}}, \& {Taibi}}]{2023MNRAS.519.1545C}
{Cardona-Barrero}, S., {Di Cintio}, A., {Battaglia}, G., {Macci{\`o}}, A.~V., \& {Taibi}, S. 2023, \mnras, 519, 1545, \dodoi{10.1093/mnras/stac3243}

\bibitem[{Cardona-Barrero {et~al.}(2020)Cardona-Barrero, Di~Cintio, Brook, Ruiz-Lara, Beasley, Falc{\'o}n-Barroso, \& Macci{\`o}}]{cardona2020nihao}
Cardona-Barrero, S., Di~Cintio, A., Brook, C.~B., {et~al.} 2020, Monthly Notices of the Royal Astronomical Society, 497, 4282

\bibitem[{Catinella {et~al.}(2010)Catinella, Schiminovich, Kauffmann, Fabello, Wang, Hummels, Lemonias, Moran, Wu, Giovanelli, {et~al.}}]{catinella2010galex}
Catinella, B., Schiminovich, D., Kauffmann, G., {et~al.} 2010, Monthly Notices of the Royal Astronomical Society, 403, 683

\bibitem[{{Chilingarian} {et~al.}(2019){Chilingarian}, {Afanasiev}, {Grishin}, {Fabricant}, \& {Moran}}]{2019ApJ...884...79C}
{Chilingarian}, I.~V., {Afanasiev}, A.~V., {Grishin}, K.~A., {Fabricant}, D., \& {Moran}, S. 2019, \apj, 884, 79, \dodoi{10.3847/1538-4357/ab4205}

\bibitem[{Conselice(2018)}]{Conselice_2018}
Conselice, C.~J. 2018, Research Notes of the AAS, 2, 43, \dodoi{10.3847/2515-5172/aab7f6}

\bibitem[{Crnojevi{\'c} {et~al.}(2016)Crnojevi{\'c}, Sand, Spekkens, Caldwell, Guhathakurta, McLeod, Seth, Simon, Strader, \& Toloba}]{crnojevic2016extended}
Crnojevi{\'c}, D., Sand, D., Spekkens, K., {et~al.} 2016, The Astrophysical Journal, 823, 19

\bibitem[{Dey {et~al.}(2019)Dey, Schlegel, Lang, Blum, Burleigh, Fan, Findlay, Finkbeiner, Herrera, Juneau, {et~al.}}]{dey2019overview}
Dey, A., Schlegel, D.~J., Lang, D., {et~al.} 2019, The Astronomical Journal, 157, 168

\bibitem[{Di~Cintio {et~al.}(2017)Di~Cintio, Brook, Dutton, Macci{\`o}, Obreja, \& Dekel}]{di2017nihao}
Di~Cintio, A., Brook, C.~B., Dutton, A.~A., {et~al.} 2017, Monthly Notices of the Royal Astronomical Society: Letters, 466, L1

\bibitem[{{Di Cintio} {et~al.}(2019){Di Cintio}, {Brook}, {Macci{\`o}}, {Dutton}, \& {Cardona-Barrero}}]{dicintio2019}
{Di Cintio}, A., {Brook}, C.~B., {Macci{\`o}}, A.~V., {Dutton}, A.~A., \& {Cardona-Barrero}, S. 2019, \mnras, 486, 2535, \dodoi{10.1093/mnras/stz985}

\bibitem[{{Dutton} {et~al.}(2007){Dutton}, {van den Bosch}, {Dekel}, \& {Courteau}}]{Dutton2007}
{Dutton}, A.~A., {van den Bosch}, F.~C., {Dekel}, A., \& {Courteau}, S. 2007, \apj, 654, 27, \dodoi{10.1086/509314}

\bibitem[{Ferr{\'e}-Mateu {et~al.}(2018)Ferr{\'e}-Mateu, Alabi, Forbes, Romanowsky, Brodie, Pandya, Mart{\'\i}n-Navarro, Bellstedt, Wasserman, Stone, {et~al.}}]{ferre2018origins}
Ferr{\'e}-Mateu, A., Alabi, A., Forbes, D.~A., {et~al.} 2018, Monthly Notices of the Royal Astronomical Society, 479, 4891

\bibitem[{Forbes {et~al.}(2020)Forbes, Alabi, Romanowsky, Brodie, \& Arimoto}]{forbes2020globular}
Forbes, D.~A., Alabi, A., Romanowsky, A.~J., Brodie, J.~P., \& Arimoto, N. 2020, Monthly Notices of the Royal Astronomical Society, 492, 4874

\bibitem[{Garrison-Kimmel {et~al.}(2019)Garrison-Kimmel, Hopkins, Wetzel, Bullock, Boylan-Kolchin, Kere{\v{s}}, Faucher-Gigu{\`e}re, El-Badry, Lamberts, Quataert, {et~al.}}]{garrison2019local}
Garrison-Kimmel, S., Hopkins, P.~F., Wetzel, A., {et~al.} 2019, Monthly Notices of the Royal Astronomical Society, 487, 1380

\bibitem[{{Gil de Paz} {et~al.}(2007){Gil de Paz}, {Boissier}, {Madore}, {Seibert}, {Joe}, {Boselli}, {Wyder}, {Thilker}, {Bianchi}, {Rey}, {Rich}, {Barlow}, {Conrow}, {Forster}, {Friedman}, {Martin}, {Morrissey}, {Neff}, {Schiminovich}, {Small}, {Donas}, {Heckman}, {Lee}, {Milliard}, {Szalay}, \& {Yi}}]{2007ApJS..173..185G}
{Gil de Paz}, A., {Boissier}, S., {Madore}, B.~F., {et~al.} 2007, \apjs, 173, 185, \dodoi{10.1086/516636}

\bibitem[{Giovanelli {et~al.}(2005)Giovanelli, Haynes, Kent, Perillat, Saintonge, Brosch, Catinella, Hoffman, Stierwalt, Spekkens, {et~al.}}]{giovanelli2005arecibo}
Giovanelli, R., Haynes, M.~P., Kent, B.~R., {et~al.} 2005, The astronomical journal, 130, 2598

\bibitem[{{Governato} {et~al.}(2015){Governato}, {Weisz}, {Pontzen}, {Loebman}, {Reed}, {Brooks}, {Behroozi}, {Christensen}, {Madau}, {Mayer}, {Shen}, {Walker}, {Quinn}, {Keller}, \& {Wadsley}}]{2015MNRAS.448..792G}
{Governato}, F., {Weisz}, D., {Pontzen}, A., {et~al.} 2015, \mnras, 448, 792, \dodoi{10.1093/mnras/stu2720}

\bibitem[{{Greco} {et~al.}(2018){Greco}, {Greene}, {Strauss}, {Macarthur}, {Flowers}, {Goulding}, {Huang}, {Kim}, {Komiyama}, {Leauthaud}, {Leisman}, {Lupton}, {Sif{\'o}n}, \& {Wang}}]{2018ApJ...857..104G}
{Greco}, J.~P., {Greene}, J.~E., {Strauss}, M.~A., {et~al.} 2018, \apj, 857, 104, \dodoi{10.3847/1538-4357/aab842}

\bibitem[{{Haardt} \& {Madau}(2012)}]{2012ApJ...746..125H}
{Haardt}, F., \& {Madau}, P. 2012, \apj, 746, 125, \dodoi{10.1088/0004-637X/746/2/125}

\bibitem[{Haynes {et~al.}(2011)Haynes, Giovanelli, Martin, Hess, Saintonge, Adams, Hallenbeck, Hoffman, Huang, Kent, {et~al.}}]{haynes2011arecibo}
Haynes, M.~P., Giovanelli, R., Martin, A.~M., {et~al.} 2011, The astronomical journal, 142, 170

\bibitem[{{Huang} {et~al.}(2014){Huang}, {Haynes}, {Giovanelli}, {Hallenbeck}, {Jones}, {Adams}, {Brinchmann}, {Chengalur}, {Hunt}, {Masters}, {Matsushita}, {Saintonge}, \& {Spekkens}}]{2014ApJ...793...40H}
{Huang}, S., {Haynes}, M.~P., {Giovanelli}, R., {et~al.} 2014, \apj, 793, 40, \dodoi{10.1088/0004-637X/793/1/40}

\bibitem[{{Iglesias-P{\'a}ramo} {et~al.}(2006){Iglesias-P{\'a}ramo}, {Buat}, {Takeuchi}, {Xu}, {Boissier}, {Boselli}, {Burgarella}, {Madore}, {Gil de Paz}, {Bianchi}, {Barlow}, {Byun}, {Donas}, {Forster}, {Friedman}, {Heckman}, {Jelinski}, {Lee}, {Malina}, {Martin}, {Milliard}, {Morrissey}, {Neff}, {Rich}, {Schiminovich}, {Seibert}, {Siegmund}, {Small}, {Szalay}, {Welsh}, \& {Wyder}}]{2006ApJS..164...38I}
{Iglesias-P{\'a}ramo}, J., {Buat}, V., {Takeuchi}, T.~T., {et~al.} 2006, \apjs, 164, 38, \dodoi{10.1086/502628}

\bibitem[{Impey {et~al.}(1988)Impey, Bothun, \& Malin}]{impey1988virgo}
Impey, C., Bothun, G., \& Malin, D. 1988, Astrophysical Journal, Part 1 (ISSN 0004-637X), vol. 330, July 15, 1988, p. 634-660., 330, 634

\bibitem[{Janowiecki {et~al.}(2019)Janowiecki, Jones, Leisman, \& Webb}]{janowiecki2019environment}
Janowiecki, S., Jones, M.~G., Leisman, L., \& Webb, A. 2019, Monthly Notices of the Royal Astronomical Society, 490, 566

\bibitem[{Javanmardi {et~al.}(2016)Javanmardi, Martinez-Delgado, Kroupa, Henkel, Crawford, Teuwen, Gabany, Hanson, Chonis, \& Neyer}]{javanmardi2016dgsat}
Javanmardi, B., Martinez-Delgado, D., Kroupa, P., {et~al.} 2016, Astronomy \& Astrophysics, 588, A89

\bibitem[{Kado-Fong {et~al.}(2022)Kado-Fong, Greene, Huang, \& Goulding}]{kado2022ultra}
Kado-Fong, E., Greene, J.~E., Huang, S., \& Goulding, A. 2022, The Astrophysical Journal, 941, 11

\bibitem[{{Kado-Fong} {et~al.}(2021){Kado-Fong}, {Petrescu}, {Mohammad}, {Greco}, {Greene}, {Adams}, {Huang}, {Leisman}, {Munshi}, {Tanoglidis}, \& {Van Nest}}]{2021ApJ...920...72K}
{Kado-Fong}, E., {Petrescu}, M., {Mohammad}, M., {et~al.} 2021, \apj, 920, 72, \dodoi{10.3847/1538-4357/ac15f0}

\bibitem[{{Kadowaki} {et~al.}(2021){Kadowaki}, {Zaritsky}, {Donnerstein}, {RS}, {Karunakaran}, \& {Spekkens}}]{2021ApJ...923..257K}
{Kadowaki}, J., {Zaritsky}, D., {Donnerstein}, R.~L., {et~al.} 2021, \apj, 923, 257, \dodoi{10.3847/1538-4357/ac2948}

\bibitem[{Karunakaran {et~al.}(2024)Karunakaran, {Motiwala}, {Spekkens}, {Zaritsky}, {Donnerstein}, \& {Dey}}]{karunakaran2024}
Karunakaran, A., {Motiwala}, K., {Spekkens}, K., {et~al.} 2024, arXiv e-prints, arXiv:2408.07119, \dodoi{10.48550/arXiv.2408.07119}

\bibitem[{Karunakaran {et~al.}(2020)Karunakaran, Spekkens, Zaritsky, Donnerstein, Kadowaki, \& Dey}]{karunakaran2020systematically}
Karunakaran, A., Spekkens, K., Zaritsky, D., {et~al.} 2020, The Astrophysical Journal, 902, 39

\bibitem[{Kauffmann {et~al.}(1993)Kauffmann, White, \& Guiderdoni}]{kauffmann1993formation}
Kauffmann, G., White, S.~D., \& Guiderdoni, B. 1993, Monthly Notices of the Royal Astronomical Society, 264, 201

\bibitem[{Keller {et~al.}(2014)Keller, Wadsley, Benincasa, \& Couchman}]{keller2014superbubble}
Keller, B., Wadsley, J., Benincasa, S., \& Couchman, H. 2014, Monthly Notices of the Royal Astronomical Society, 442, 3013

\bibitem[{{Kelly}(2007)}]{2007ApJ...665.1489K}
{Kelly}, B.~C. 2007, \apj, 665, 1489, \dodoi{10.1086/519947}

\bibitem[{{Kennicutt}(1998)}]{Kennicutt1998}
{Kennicutt}, Robert~C., J. 1998, \apj, 498, 541, \dodoi{10.1086/305588}

\bibitem[{{Kennicutt} {et~al.}(1994){Kennicutt}, {Tamblyn}, \& {Congdon}}]{1994ApJ...435...22K}
{Kennicutt}, Robert~C., J., {Tamblyn}, P., \& {Congdon}, C.~E. 1994, \apj, 435, 22, \dodoi{10.1086/174790}

\bibitem[{{Koda} {et~al.}(2015){Koda}, {Yagi}, {Yamanoi}, \& {Komiyama}}]{2015ApJ...807L...2K}
{Koda}, J., {Yagi}, M., {Yamanoi}, H., \& {Komiyama}, Y. 2015, \apjl, 807, L2, \dodoi{10.1088/2041-8205/807/1/L2}

\bibitem[{{Kormendy} \& {Sanders}(1992)}]{1992ApJ...390L..53K}
{Kormendy}, J., \& {Sanders}, D.~B. 1992, \apjl, 390, L53, \dodoi{10.1086/186370}

\bibitem[{Kov{\'a}cs {et~al.}(2019)Kov{\'a}cs, Bogd{\'a}n, \& Canning}]{kovacs2019constraining}
Kov{\'a}cs, O.~E., Bogd{\'a}n, {\'A}., \& Canning, R.~E. 2019, The Astrophysical Journal Letters, 879, L12

\bibitem[{Kroupa(2001)}]{kroupa2001variation}
Kroupa, P. 2001, Monthly Notices of the Royal Astronomical Society, 322, 231

\bibitem[{Lange {et~al.}(2015)Lange, Driver, Robotham, Kelvin, Graham, Alpaslan, Andrews, Baldry, Bamford, Bland-Hawthorn, {et~al.}}]{lange2015galaxy}
Lange, R., Driver, S.~P., Robotham, A.~S., {et~al.} 2015, Monthly Notices of the Royal Astronomical Society, 447, 2603

\bibitem[{Leisman {et~al.}(2017)Leisman, Haynes, Janowiecki, Hallenbeck, J{\'o}zsa, Giovanelli, Adams, Neira, Cannon, Janesh, {et~al.}}]{leisman2017almost}
Leisman, L., Haynes, M.~P., Janowiecki, S., {et~al.} 2017, The Astrophysical Journal, 842, 133

\bibitem[{Lemos {et~al.}(2024)Lemos, Sharief, Malkin, Perreault-Levasseur, \& Hezaveh}]{lemos2024pqmass}
Lemos, P., Sharief, S., Malkin, N., Perreault-Levasseur, L., \& Hezaveh, Y. 2024, arXiv preprint arXiv:2402.04355

\bibitem[{{Macci{\`o}} {et~al.}(2020){Macci{\`o}}, {Crespi}, {Blank}, \& {Kang}}]{Maccio2020}
{Macci{\`o}}, A.~V., {Crespi}, S., {Blank}, M., \& {Kang}, X. 2020, \mnras, 495, L46, \dodoi{10.1093/mnrasl/slaa058}

\bibitem[{{Macci{\`o}} {et~al.}(2016){Macci{\`o}}, {Udrescu}, {Dutton}, {Obreja}, {Wang}, {Stinson}, \& {Kang}}]{Maccio2016}
{Macci{\`o}}, A.~V., {Udrescu}, S.~M., {Dutton}, A.~A., {et~al.} 2016, \mnras, 463, L69, \dodoi{10.1093/mnrasl/slw147}

\bibitem[{{Mart{\'\i}n-Navarro} {et~al.}(2019){Mart{\'\i}n-Navarro}, {Romanowsky}, {Brodie}, {Ferr{\'e}-Mateu}, {Alabi}, {Forbes}, {Sharina}, {Villaume}, {Pandya}, \& {Martinez-Delgado}}]{2019MNRAS.484.3425M}
{Mart{\'\i}n-Navarro}, I., {Romanowsky}, A.~J., {Brodie}, J.~P., {et~al.} 2019, \mnras, 484, 3425, \dodoi{10.1093/mnras/stz252}

\bibitem[{{Mart{\'\i}nez-Delgado} {et~al.}(2016){Mart{\'\i}nez-Delgado}, {L{\"a}sker}, {Sharina}, {Toloba}, {Fliri}, {Beaton}, {Valls-Gabaud}, {Karachentsev}, {Chonis}, {Grebel}, {Forbes}, {Romanowsky}, {Gallego-Laborda}, {Teuwen}, {G{\'o}mez-Flechoso}, {Wang}, {Guhathakurta}, {Kaisin}, \& {Ho}}]{2016AJ....151...96M}
{Mart{\'\i}nez-Delgado}, D., {L{\"a}sker}, R., {Sharina}, M., {et~al.} 2016, \aj, 151, 96, \dodoi{10.3847/0004-6256/151/4/96}

\bibitem[{{Menon} {et~al.}(2015){Menon}, {Wesolowski}, {Zheng}, {Jetley}, {Kale}, {Quinn}, \& {Governato}}]{Menon2015}
{Menon}, H., {Wesolowski}, L., {Zheng}, G., {et~al.} 2015, Computational Astrophysics and Cosmology, 2, 1, \dodoi{10.1186/s40668-015-0007-9}

\bibitem[{{Merritt} {et~al.}(2016){Merritt}, {van Dokkum}, {Danieli}, {Abraham}, {Zhang}, {Karachentsev}, \& {Makarova}}]{2016ApJ...833..168M}
{Merritt}, A., {van Dokkum}, P., {Danieli}, S., {et~al.} 2016, \apj, 833, 168, \dodoi{10.3847/1538-4357/833/2/168}

\bibitem[{Mihos {et~al.}(2015)Mihos, Durrell, Ferrarese, Feldmeier, C{\^o}t{\'e}, Peng, Harding, Liu, Gwyn, \& Cuillandre}]{mihos2015galaxies}
Mihos, J.~C., Durrell, P.~R., Ferrarese, L., {et~al.} 2015, The Astrophysical Journal Letters, 809, L21

\bibitem[{Moreno {et~al.}(2022)Moreno, Danieli, Bullock, Feldmann, Hopkins, Catmabacak, Gurvich, Lazar, Klein, Hummels, {et~al.}}]{moreno2022galaxies}
Moreno, J., Danieli, S., Bullock, J.~S., {et~al.} 2022, Nature Astronomy, 6, 496

\bibitem[{{M{\"u}ller} {et~al.}(2018){M{\"u}ller}, {Jerjen}, \& {Binggeli}}]{2018A&A...615A.105M}
{M{\"u}ller}, O., {Jerjen}, H., \& {Binggeli}, B. 2018, \aap, 615, A105, \dodoi{10.1051/0004-6361/201832897}

\bibitem[{{Nagesh} {et~al.}(2024){Nagesh}, {Freundlich}, {Famaey}, {B{\'\i}lek}, {Candlish}, {Ibata}, \& {M{\"u}ller}}]{2024A&A...690A.149N}
{Nagesh}, S.~T., {Freundlich}, J., {Famaey}, B., {et~al.} 2024, \aap, 690, A149, \dodoi{10.1051/0004-6361/202450757}

\bibitem[{Papastergis {et~al.}(2017)Papastergis, Adams, \& Romanowsky}]{papastergis2017hi}
Papastergis, E., Adams, E., \& Romanowsky, A. 2017, Astronomy \& Astrophysics, 601, L10

\bibitem[{{Peng} {et~al.}(2010){Peng}, {Ho}, {Impey}, \& {Rix}}]{2010AJ....139.2097P}
{Peng}, C.~Y., {Ho}, L.~C., {Impey}, C.~D., \& {Rix}, H.-W. 2010, \aj, 139, 2097, \dodoi{10.1088/0004-6256/139/6/2097}

\bibitem[{{Planck Collaboration} {et~al.}(2014){Planck Collaboration}, {Ade}, {Aghanim}, {Armitage-Caplan}, {Arnaud}, {Ashdown}, {Atrio-Barandela}, {Aumont}, {Baccigalupi}, {Banday}, \& et~al.}]{planck14}
{Planck Collaboration}, {Ade}, P.~A.~R., {Aghanim}, N., {et~al.} 2014, \aap, 571, A16, \dodoi{10.1051/0004-6361/201321591}

\bibitem[{{Pontzen} {et~al.}(2013){Pontzen}, {Ro{\v{s}}kar}, {Stinson}, \& {Woods}}]{Pontzen2013}
{Pontzen}, A., {Ro{\v{s}}kar}, R., {Stinson}, G., \& {Woods}, R. 2013, {pynbody: N-Body/SPH analysis for python}.
\newblock \doeprint{1305.002}

\bibitem[{{Prole} {et~al.}(2018){Prole}, {Davies}, {Keenan}, \& {Davies}}]{2018MNRAS.478..667P}
{Prole}, D.~J., {Davies}, J.~I., {Keenan}, O.~C., \& {Davies}, L.~J.~M. 2018, \mnras, 478, 667, \dodoi{10.1093/mnras/sty1021}

\bibitem[{{Prole} {et~al.}(2019){Prole}, {van der Burg}, {Hilker}, \& {Davies}}]{2019MNRAS.488.2143P}
{Prole}, D.~J., {van der Burg}, R.~F.~J., {Hilker}, M., \& {Davies}, J.~I. 2019, \mnras, 488, 2143, \dodoi{10.1093/mnras/stz1843}

\bibitem[{{Rahmati} {et~al.}(2013){Rahmati}, {Pawlik}, {Rai{\v{c}}evi{\'c}}, \& {Schaye}}]{2013MNRAS.430.2427R}
{Rahmati}, A., {Pawlik}, A.~H., {Rai{\v{c}}evi{\'c}}, M., \& {Schaye}, J. 2013, \mnras, 430, 2427, \dodoi{10.1093/mnras/stt066}

\bibitem[{Rom{\'a}n \& Trujillo(2017)}]{roman2017ultra}
Rom{\'a}n, J., \& Trujillo, I. 2017, Monthly Notices of the Royal Astronomical Society, 468, 4039

\bibitem[{Rong {et~al.}(2017)Rong, Guo, Gao, Liao, Xie, Puzia, Sun, \& Pan}]{rong2017universe}
Rong, Y., Guo, Q., Gao, L., {et~al.} 2017, Monthly Notices of the Royal Astronomical Society, 470, 4231

\bibitem[{{Saintonge} \& {Catinella}(2022)}]{saintonge22}
{Saintonge}, A., \& {Catinella}, B. 2022, \araa, 60, 319, \dodoi{10.1146/annurev-astro-021022-043545}

\bibitem[{Santos-Santos {et~al.}(2018)Santos-Santos, Di~Cintio, Brook, Macci{\`o}, Dutton, \& Dom{\'\i}nguez-Tenreiro}]{santos2018nihao}
Santos-Santos, I.~M., Di~Cintio, A., Brook, C.~B., {et~al.} 2018, Monthly Notices of the Royal Astronomical Society, 473, 4392

\bibitem[{{Schechter}(1976)}]{1976ApJ...203..297S}
{Schechter}, P. 1976, \apj, 203, 297, \dodoi{10.1086/154079}

\bibitem[{{Schombert} {et~al.}(2011){Schombert}, {Maciel}, \& {McGaugh}}]{2011AdAst2011E..12S}
{Schombert}, J., {Maciel}, T., \& {McGaugh}, S. 2011, Advances in Astronomy, 2011, 143698, \dodoi{10.1155/2011/143698}

\bibitem[{Shen {et~al.}(2010)Shen, Wadsley, \& Stinson}]{shen2010enrichment}
Shen, S., Wadsley, J., \& Stinson, G. 2010, Monthly Notices of the Royal Astronomical Society, 407, 1581

\bibitem[{{Spekkens} \& {Karunakaran}(2018)}]{2018ApJ...855...28S}
{Spekkens}, K., \& {Karunakaran}, A. 2018, \apj, 855, 28, \dodoi{10.3847/1538-4357/aa94be}

\bibitem[{{Stinson} {et~al.}(2006){Stinson}, {Seth}, {Katz}, {Wadsley}, {Governato}, \& {Quinn}}]{Stinson2006}
{Stinson}, G., {Seth}, A., {Katz}, N., {et~al.} 2006, \mnras, 373, 1074, \dodoi{10.1111/j.1365-2966.2006.11097.x}

\bibitem[{Stinson {et~al.}(2006)Stinson, Seth, Katz, Wadsley, Governato, \& Quinn}]{stinson2006star}
Stinson, G., Seth, A., Katz, N., {et~al.} 2006, Monthly Notices of the Royal Astronomical Society, 373, 1074

\bibitem[{{Stinson} {et~al.}(2013){Stinson}, {Brook}, {Macci{\`o}}, {Wadsley}, {Quinn}, \& {Couchman}}]{Stinson2013}
{Stinson}, G.~S., {Brook}, C., {Macci{\`o}}, A.~V., {et~al.} 2013, \mnras, 428, 129, \dodoi{10.1093/mnras/sts028}

\bibitem[{{Sung} {et~al.}(2002){Sung}, {Chun}, {Freeman}, \& {Chaboyer}}]{2002ASPC..273..341S}
{Sung}, E.-C., {Chun}, M.-S., {Freeman}, K.~C., \& {Chaboyer}, B. 2002, in Astronomical Society of the Pacific Conference Series, Vol. 273, The Dynamics, Structure \& History of Galaxies: A Workshop in Honour of Professor Ken Freeman, ed. G.~S. {Da Costa}, E.~M. {Sadler}, \& H.~{Jerjen}, 341

\bibitem[{{Toloba} {et~al.}(2016){Toloba}, {Sand}, {Guhathakurta}, {Chiboucas}, {Crnojevi{\'c}}, \& {Simon}}]{2016ApJ...830L..21T}
{Toloba}, E., {Sand}, D., {Guhathakurta}, P., {et~al.} 2016, \apjl, 830, L21, \dodoi{10.3847/2041-8205/830/1/L21}

\bibitem[{{Tremmel} {et~al.}(2015){Tremmel}, {Governato}, {Volonteri}, \& {Quinn}}]{2015MNRAS.451.1868T}
{Tremmel}, M., {Governato}, F., {Volonteri}, M., \& {Quinn}, T.~R. 2015, \mnras, 451, 1868, \dodoi{10.1093/mnras/stv1060}

\bibitem[{{Tremmel} {et~al.}(2017){Tremmel}, {Karcher}, {Governato}, {Volonteri}, {Quinn}, {Pontzen}, {Anderson}, \& {Bellovary}}]{2017MNRAS.470.1121T}
{Tremmel}, M., {Karcher}, M., {Governato}, F., {et~al.} 2017, \mnras, 470, 1121, \dodoi{10.1093/mnras/stx1160}

\bibitem[{Tremmel {et~al.}(2020)Tremmel, Wright, Brooks, Munshi, Nagai, \& Quinn}]{tremmel2020formation}
Tremmel, M., Wright, A.~C., Brooks, A.~M., {et~al.} 2020, Monthly Notices of the Royal Astronomical Society, 497, 2786

\bibitem[{Tremmel {et~al.}(2019)Tremmel, Quinn, Ricarte, Babul, Chadayammuri, Natarajan, Nagai, Pontzen, \& Volonteri}]{tremmel2019introducing}
Tremmel, M., Quinn, T.~R., Ricarte, A., {et~al.} 2019, Monthly Notices of the Royal Astronomical Society, 483, 3336

\bibitem[{Trujillo {et~al.}(2017)Trujillo, Roman, Almeida, {et~al.}}]{trujillo2017nearest}
Trujillo, I., Roman, J., Almeida, J.~S., {et~al.} 2017, arXiv preprint arXiv:1701.03804

\bibitem[{van Der~Burg {et~al.}(2016)van Der~Burg, Muzzin, \& Hoekstra}]{van2016abundance}
van Der~Burg, R.~F., Muzzin, A., \& Hoekstra, H. 2016, Astronomy \& Astrophysics, 590, A20

\bibitem[{van Der~Burg {et~al.}(2017)van Der~Burg, Hoekstra, Muzzin, Sif{\'o}n, Viola, Bremer, Brough, Driver, Erben, Heymans, {et~al.}}]{van2017abundance}
van Der~Burg, R.~F., Hoekstra, H., Muzzin, A., {et~al.} 2017, Astronomy \& Astrophysics, 607, A79

\bibitem[{Van~Dokkum {et~al.}(2018)Van~Dokkum, Danieli, Cohen, Merritt, Romanowsky, Abraham, Brodie, Conroy, Lokhorst, Mowla, {et~al.}}]{van2018galaxy}
Van~Dokkum, P., Danieli, S., Cohen, Y., {et~al.} 2018, Nature, 555, 629

\bibitem[{van Dokkum {et~al.}(2015{\natexlab{a}})van Dokkum, Abraham, Merritt, Zhang, Geha, \& Conroy}]{van2015forty}
van Dokkum, P.~G., Abraham, R., Merritt, A., {et~al.} 2015{\natexlab{a}}, The Astrophysical Journal Letters, 798, L45

\bibitem[{Van~Dokkum {et~al.}(2013)Van~Dokkum, Leja, Nelson, Patel, Skelton, Momcheva, Brammer, Whitaker, Lundgren, Fumagalli, {et~al.}}]{van2013assembly}
Van~Dokkum, P.~G., Leja, J., Nelson, E.~J., {et~al.} 2013, The Astrophysical Journal Letters, 771, L35

\bibitem[{van Dokkum {et~al.}(2015{\natexlab{b}})van Dokkum, Romanowsky, Abraham, Brodie, Conroy, Geha, Merritt, Villaume, \& Zhang}]{van2015spectroscopic}
van Dokkum, P.~G., Romanowsky, A.~J., Abraham, R., {et~al.} 2015{\natexlab{b}}, The Astrophysical Journal Letters, 804, L26

\bibitem[{{Van Nest} {et~al.}(2022){Van Nest}, {Munshi}, {Wright}, {Tremmel}, {Brooks}, {Nagai}, \& {Quinn}}]{2022ApJ...926...92V}
{Van Nest}, J.~D., {Munshi}, F., {Wright}, A.~C., {et~al.} 2022, \apj, 926, 92, \dodoi{10.3847/1538-4357/ac43b7}

\bibitem[{Venhola {et~al.}(2017)Venhola, Peletier, Laurikainen, Salo, Lisker, Iodice, Capaccioli, Kleijn, Valentijn, Mieske, {et~al.}}]{venhola2017fornax}
Venhola, A., Peletier, R., Laurikainen, E., {et~al.} 2017, Astronomy \& Astrophysics, 608, A142

\bibitem[{Vogelsberger {et~al.}(2020)Vogelsberger, Marinacci, Torrey, \& Puchwein}]{vogelsberger2020cosmological}
Vogelsberger, M., Marinacci, F., Torrey, P., \& Puchwein, E. 2020, Nature Reviews Physics, 2, 42

\bibitem[{{Wadsley} {et~al.}(2017){Wadsley}, {Keller}, \& {Quinn}}]{Wadsley2017}
{Wadsley}, J.~W., {Keller}, B.~W., \& {Quinn}, T.~R. 2017, \mnras, 471, 2357, \dodoi{10.1093/mnras/stx1643}

\bibitem[{Wadsley {et~al.}(2004)Wadsley, Stadel, \& Quinn}]{wadsley2004gasoline}
Wadsley, J.~W., Stadel, J., \& Quinn, T. 2004, New astronomy, 9, 137

\bibitem[{Wang {et~al.}(2015)Wang, Dutton, Stinson, Macci{\`o}, Penzo, Kang, Keller, \& Wadsley}]{wang2015nihao}
Wang, L., Dutton, A.~A., Stinson, G.~S., {et~al.} 2015, Monthly Notices of the Royal Astronomical Society, 454, 83

\bibitem[{White \& Rees(1978)}]{white1978core}
White, S.~D., \& Rees, M.~J. 1978, Monthly Notices of the Royal Astronomical Society, 183, 341

\bibitem[{{White} \& {Frenk}(1991)}]{1991ApJ...379...52W}
{White}, S. D.~M., \& {Frenk}, C.~S. 1991, \apj, 379, 52, \dodoi{10.1086/170483}

\bibitem[{Wright {et~al.}(2021)Wright, Tremmel, Brooks, Munshi, Nagai, Sharma, \& Quinn}]{wright2021formation}
Wright, A.~C., Tremmel, M., Brooks, A.~M., {et~al.} 2021, Monthly Notices of the Royal Astronomical Society, 502, 5370

\bibitem[{{York} {et~al.}(2000){York}, {Adelman}, {Anderson}, {Anderson}, {Annis}, {Bahcall}, {Bakken}, {Barkhouser}, {Bastian}, {Berman}, {Boroski}, {Bracker}, {Briegel}, {Briggs}, {Brinkmann}, {Brunner}, {Burles}, {Carey}, {Carr}, {Castander}, {Chen}, {Colestock}, {Connolly}, {Crocker}, {Csabai}, {Czarapata}, {Davis}, {Doi}, {Dombeck}, {Eisenstein}, {Ellman}, {Elms}, {Evans}, {Fan}, {Federwitz}, {Fiscelli}, {Friedman}, {Frieman}, {Fukugita}, {Gillespie}, {Gunn}, {Gurbani}, {de Haas}, {Haldeman}, {Harris}, {Hayes}, {Heckman}, {Hennessy}, {Hindsley}, {Holm}, {Holmgren}, {Huang}, {Hull}, {Husby}, {Ichikawa}, {Ichikawa}, {Ivezi{\'c}}, {Kent}, {Kim}, {Kinney}, {Klaene}, {Kleinman}, {Kleinman}, {Knapp}, {Korienek}, {Kron}, {Kunszt}, {Lamb}, {Lee}, {Leger}, {Limmongkol}, {Lindenmeyer}, {Long}, {Loomis}, {Loveday}, {Lucinio}, {Lupton}, {MacKinnon}, {Mannery}, {Mantsch}, {Margon}, {McGehee}, {McKay}, {Meiksin}, {Merelli}, {Monet}, {Munn}, {Narayanan}, {Nash}, {Neilsen}, {Neswold}, {Newberg}, {Nichol}, {Nicinski},
  {Nonino}, {Okada}, {Okamura}, {Ostriker}, {Owen}, {Pauls}, {Peoples}, {Peterson}, {Petravick}, {Pier}, {Pope}, {Pordes}, {Prosapio}, {Rechenmacher}, {Quinn}, {Richards}, {Richmond}, {Rivetta}, {Rockosi}, {Ruthmansdorfer}, {Sandford}, {Schlegel}, {Schneider}, {Sekiguchi}, {Sergey}, {Shimasaku}, {Siegmund}, {Smee}, {Smith}, {Snedden}, {Stone}, {Stoughton}, {Strauss}, {Stubbs}, {SubbaRao}, {Szalay}, {Szapudi}, {Szokoly}, {Thakar}, {Tremonti}, {Tucker}, {Uomoto}, {Vanden Berk}, {Vogeley}, {Waddell}, {Wang}, {Watanabe}, {Weinberg}, {Yanny}, {Yasuda}, \& {SDSS Collaboration}}]{2000AJ....120.1579Y}
{York}, D.~G., {Adelman}, J., {Anderson}, John~E., J., {et~al.} 2000, \aj, 120, 1579, \dodoi{10.1086/301513}

\bibitem[{{Yozin} \& {Bekki}(2015)}]{2015MNRAS.452..937Y}
{Yozin}, C., \& {Bekki}, K. 2015, \mnras, 452, 937, \dodoi{10.1093/mnras/stv1073}

\bibitem[{{Zaritsky} {et~al.}(2023){Zaritsky}, {Donnerstein}, {Dey}, {Karunakaran}, {Kadowaki}, {Khim}, {Spekkens}, \& {Zhang}}]{2023ApJS..267...27Z}
{Zaritsky}, D., {Donnerstein}, R., {Dey}, A., {et~al.} 2023, \apjs, 267, 27, \dodoi{10.3847/1538-4365/acdd71}

\bibitem[{{Zaritsky} {et~al.}(2021){Zaritsky}, Donnerstein, Karunakaran, Barbosa, Dey, Kadowaki, Spekkens, \& Zhang}]{zaritsky2021systematically}
{Zaritsky}, D., Donnerstein, R., Karunakaran, A., {et~al.} 2021, The Astrophysical Journal Supplement Series, 257, 60

\bibitem[{{Zaritsky} {et~al.}(2022){Zaritsky}, Donnerstein, Karunakaran, Barbosa, Dey, Kadowaki, Spekkens, \& Zhang}]{zaritsky2022systematically}
---. 2022, The Astrophysical Journal Supplement Series, 261, 11

\bibitem[{{Zaritsky} {et~al.}(2018){Zaritsky}, Donnerstein, Dey, Kadowaki, Zhang, Karunakaran, Mart{\'\i}nez-Delgado, Rahman, \& Spekkens}]{zaritsky2018systematically}
{Zaritsky}, D., Donnerstein, R., Dey, A., {et~al.} 2018, The Astrophysical Journal Supplement Series, 240, 1

\bibitem[{{Zaritsky} {et~al.}(2019){Zaritsky}, {Donnerstein}, {Dey}, {Kadowaki}, {Zhang}, {Karunakaran}, {Mart{\'\i}nez-Delgado}, {Rahman}, \& {Spekkens}}]{2019ApJS..240....1Z}
{Zaritsky}, D., {Donnerstein}, R., {Dey}, A., {et~al.} 2019, \apjs, 240, 1, \dodoi{10.3847/1538-4365/aaefe9}

\bibitem[{{Zhang} {et~al.}(2017){Zhang}, {Puzia}, \& {Weisz}}]{2017ApJS..233...13Z}
{Zhang}, H.-X., {Puzia}, T.~H., \& {Weisz}, D.~R. 2017, \apjs, 233, 13, \dodoi{10.3847/1538-4365/aa937b}

\bibitem[{{Zhang} {et~al.}(2024){Zhang}, {Bi}, \& {Yin}}]{2024arXiv240311403Z}
{Zhang}, Z.-C., {Bi}, X.-J., \& {Yin}, P.-F. 2024, arXiv e-prints, arXiv:2403.11403, \dodoi{10.48550/arXiv.2403.11403}

\bibitem[{Zitrin {et~al.}(2009)Zitrin, Brosch, \& Bilenko}]{zitrin2009star}
Zitrin, A., Brosch, N., \& Bilenko, B. 2009, Monthly Notices of the Royal Astronomical Society, 399, 924

\bibitem[{{Z{\"o}ller} {et~al.}(2024){Z{\"o}ller}, {Kluge}, {Staiger}, \& {Bender}}]{2024ApJS..271...52Z}
{Z{\"o}ller}, R., {Kluge}, M., {Staiger}, B., \& {Bender}, R. 2024, \apjs, 271, 52, \dodoi{10.3847/1538-4365/ad2775}

\end{thebibliography}
\bibliographystyle{aasjournal}

\end{document}